\documentclass[usenatbib, usegraphicx]{mn2e}
\usepackage{times}

\newcommand{\cm}{{\rm cm}} 
\newcommand{\yr}{{\rm yr}} 

\newcommand{\kms}{{\rm km}\,{\rm s}^{-1}} 
\newcommand{\K}{{\rm K}}
\newcommand{\pc}{{\rm pc}} 
\newcommand{\kpc}{{\rm kpc}}

\newcommand{\erg}{{\rm ergs}}
\newcommand{\Msun}{{{\rm M}_\odot}}
\newcommand{\hMpc}{h^{-1}\,{\rm Mpc}}

\newcommand{\owls}{{\sc owls}}
\newcommand{\reference}{{\sc reference}}
\newcommand{\NOSN}{{\sc nosn}}
\newcommand{\NOZCOOL}{{\sc nozcool}}
\newcommand{\NOSNNOZCOOL}{{\sc nosn\_nozcool}}
\newcommand{\wdens}{{\sc wdens}}
\newcommand{\wvcirc}{{\sc wvcirc}}
\newcommand{\wthermal}{{\sc wthermal}}
\newcommand{\dblimf}{{\sc dblimf}}

\newcommand{\AGN}{{\sc agn}}
\newcommand{\agn}{{\sc agn}}
\newcommand{\MILL}{{\sc mill}}

\title[The cosmic distribution of metals]{The effect of variations in
the input physics on the cosmic distribution of metals predicted by
simulations}

\author[R. P. C. Wiersma et al.]{%
Robert P. C. Wiersma,$^{1,2}$\thanks{E-mail: wiersma@mpa-garching.mpg.de} 
Joop Schaye,$^2$ and Tom Theuns$^{3,4}$\\ 
$^1$Max-Planck-Institut f\"ur Astrophysik,
Karl-Schwarzschild-Strasse 1, D-85478 Garching, Germany\\
$^2$Leiden Observatory, Leiden University, P.O. Box 9513, 2300 RA Leiden, the Netherlands \\
$^3$Institute for Computational Cosmology, Department of Physics, University of Durham, South Road, Durham, DH1 3LE, UK\\
$^4$Department of Physics, University of Antwerp, Groenenborgerlaan 171, B-2020 Antwerpen, Belgium
}

\begin{document}

\maketitle
\begin{abstract}
We investigate how a range of physical processes affect the cosmic
  metal distribution using a suite of cosmological, hydrodynamical simulations. 
Focusing 
  on redshifts $z=0$ and 2, we study the
  metallicities and metal mass fractions for stars as well as for the
  interstellar medium (ISM), and several more diffuse gas phases. 
We vary the cooling rates, star formation law, structure of the ISM, properties of galactic winds, feedback from AGN, supernova type Ia time delays, reionization, stellar initial mass function, and cosmology. 
In all models stars and the 
warm-hot intergalactic medium (WHIM) constitute the dominant repository
of metals, while for $z \ga 2$ the ISM is also important.  In models with
galactic winds, predictions for the metallicities of the
various phases vary at the factor of two level and are broadly
consistent with observations. The exception is the cold-warm intergalactic medium (IGM), whose
metallicity varies at the order of magnitude level if the prescription
for galactic winds is varied, even for a fixed wind energy per unit
stellar mass formed, and falls far below the observed values if winds
are not included. At the other extreme, the metallicity of the intracluster medium (ICM) is
largely insensitive to the presence of galactic winds, indicating that its
enrichment is regulated by other processes.  The mean metallicities of
stars ($\sim Z_\odot$), the ICM ($\sim 10^{-1}\,Z_\odot$), and the
WHIM ($\sim 10^{-1}\,Z_\odot$) evolve only slowly, while those of
the cold halo gas and the IGM increase by more than an order of
magnitude from $z=5$ to 0. 
Higher velocity outflows are more efficient at
transporting metals to low densities, but actually predict lower metallicities for the cold-warm IGM
since the winds shock-heat the gas to high
temperatures, thereby increasing the fraction of the metals residing
in, but not the metallicity of, the WHIM. 
Besides galactic winds driven by feedback from
star formation, the metal distribution is most sensitive to the
inclusion of metal-line cooling and feedback from AGN. We conclude
that observations of the metallicity of the low-density IGM
have the potential to constrain the poorly understood feedback
processes that are central to current models of the formation and
evolution of galaxies.
\end{abstract}


\section{Introduction}
The spatial distribution of elements synthesised in stars (henceforth `metals') provides an archaeological record of past star formation activity and of the
various energetic phenomena that stirred and mixed these metals. Recent cosmological simulations of galaxy formation follow the different stellar evolutionary channels through which metals are produced, and include some processes that cause metals to escape from their parent galaxy. Here we will investigate to what extent these simulations produce realistic enrichment patterns, which physical processes affect the metal distribution, and how robust the predictions are.

Observations can constrain integrated stellar and nebular (ISM) abundances
of galaxies (for $z\la 1$ see e.g.\ \citealt{Kobulnicky1999, Kunth2000, Teta04,
  Dunne2003}, for $z\approx 1$ see \citealt{Churchill2007},  for $z\ga 2$ see \citealt{Eeta06,Mannucci2009}). 
  Metals in cold gas can be observed in absorption against a background source, see the damped Lyman-alpha observations at $z \approx 3$ \citep[e.g.][]{Pettini1994,  Prochaska2003} and at lower \citep[e.g.][]{Meiring2009} and higher redshift
\citep[e.g.][]{Ando2007,Price2007}. If the background source is sufficiently bright, then this technique can also be applied to more diffuse gas \citep[e.g.][] {SC96, Cowie1998,
  Ellison2000, Seta00a, Seta03, Schaye2007, Simcoe2004, Scannapieco2006b, Aguirre2008}. The metallicity of the ICM is inferred from X-ray observations
\citep[e.g.][]{Mushotzky1996, dp2006, Sato2007a, Rasmussen2007, Maughan2008, Leccardi2008, Snowden2008}.  

These
observations use different tools to observe a variety of elements
in a variety of ionization states, and it is not always obvious how to
convert all these measurements to a common `metallicity'. Often
this is done assuming that the relative abundances of elements equal those measured
in the Sun. Unfortunately even the assumed metallicity of the Sun
itself varies between authors. Here we assume $Z_\odot=0.0127$, and
solar abundances from the default settings of {\sc cloudy} (version
07.02, last described by Ferland et al.\ 1998).

Convolving metallicity with the fraction of baryonic mass in each of
the different phases then yields a census of cosmic metals
\citep[e.g.][]{Fuku98}. It must, however, be kept in mind that a large
fraction of the metals are potentially not accounted for by the
current data. For example, a large portion of $z=0$ baryons are
thought to reside in the warm-hot intergalactic medium (WHIM), which
has not yet been convincingly detected. If, as is likely, the WHIM is
also metal enriched, then it could harbour a significant amount of
metals. Similarly, the prevalence of hot ($T\ga 10^5$~K) metals at
$z\ga 2$ is currently poorly constrained. Gas cooling, star formation,
galaxy interactions, ram pressure stripping, stellar and galactic
winds all cause gas and hence metals to be cycled between the
different phases, which makes it complicated to relate the observed
metal distribution to the original source and/or enrichment process.

A theoretical calculation of the census of cosmic metals must take into
account their production by nucleosynthesis, their initial
distribution, and the mixing that occurs in later
stages. 
Numerical simulations that include such \lq chemo-dynamics\rq\ have
become increasingly sophisticated since the early work by
\citet{TBH92}, usually concentrating on the evolution
of a single galaxy \citep{SM95, RVN96,
  Berczik1999,Reta01,K01a,KG03,Kobayashi2004,MU06a, Seta06a, Geta07,Bekki2009a, Rahimi2010, Pasetto2010}, or a
cluster of galaxies \citep[e.g.][]{LPC02, Valdarnini2003,Toeta04,
  Sommer-larsen2005, Reta06, Tornatore2007b, Fabjan2010} in a cosmological
context. These simulations include interpolation tables from stellar
evolution calculations for the production of metals, but implement the
enrichment processes explicitly, for example by injecting metal-enriched
gas near a site of star formation in a galactic wind. The fluxes
of metals between phases due to gas cooling and galaxy/gas
interactions are computed explicitly by these hydro-codes. Early
simulations that were also used to look at the metals outside galaxies include
\cite{Cen1999a}, \cite{Mosconi2001} and \cite{Theuns2002}. Subsequent authors
implemented more physics while increasing resolution and/or box size in
order to minimise numerical effects \citep[e.g.][]{Seta05, OD06, Cen2006,
  Brook2007, Kobayashi2007,Oppenheimer2008, Wiersma2009b,Wiersma2010,Shen2009, Tornatore2009}.

In this paper we use cosmological hydrodynamical simulations of the
formation of galaxies to attempt to answer the question `{\em Where
  are the metals}?', by computing the evolution of the fraction of metals in stars and various gas phases. Our suite of numerical simulations
(the OWLS suite; \citealt{Schaye2010}), includes runs that differ in terms of their resolution,
input physics, and numerical implementation of physical processes. We use it 
to investigate what physical processes are most important, how reliable the predictions are, and to what extent
these depend on the sometimes poorly understood physics \cite[see][for a similar invesitgation]{Sommer-Larsen2008}. In \cite{Wiersma2009b} we introduced the method and  described some of the numerical issues involved, but analysed only a single physical model (the OWLS reference model). Here we vary the cooling rates, star formation law, structure of the ISM, properties of galactic winds, feedback from AGN, supernova type Ia time delays, reionization, stellar initial mass function, and cosmology.

This paper is organised as follows. Section~\ref{sec-method}
introduces the simulations used. Those familiar with the OWLS project
may skip to the results in Section~\ref{sec-WATM}, which begins with
an overview before discussing the physical variations which are the
most relevant to the metal distribution. This section closes with a
summary of all the simulations in the OWLS suite in
Section~\ref{sec-sum}. In Section~\ref{sec-concs} we present our
conclusions.

\section{Simulations}
\label{sec-method}

The \owls\ suite \citep{Schaye2010} consists of more than fifty large, cosmological, gas-dynamical simulations
in periodic boxes, performed using the $N$-body Tree-PM, SPH code
\textsc{gadget iii}; see Tables~1 and 2 in \cite{Schaye2010} for a full list of parameters. \textsc{gadget iii} is an updated version of \textsc{gadget ii} \citep{Springel2005}, to which we added new physics modules for star formation \citep{Schaye2008}, feedback from supernovae (SNe) in the form of galactic winds \citep{Dallavecchia2008}, feedback from accreting black holes \citep{Booth2009}, radiative cooling and heating in the presence of an ionizing background \citep{Wiersma2009}, and stellar evolution \citep{Wiersma2009b}.  In this simulation suite, numerical and poorly known physical parameters are varied with respect to a \lq reference\rq\ model, to assess which conclusions are robust, and which processes dominate. The cosmological parameters of the reference model, its subgrid and their models and their their numerical implementation, are discussed briefly in the next section. {\sc GIMIC} \citep{Crain2009} is a complementary suite of simulations performed with the same simulation code, but employs a single set of parameters to investigate how star formation depends on environment, by simulating regions picked from a large volume  simulation.  Here we give a short overview of the code, concentrating especially on those processes that are directly
relevant to metal enrichment.

\subsection{The {\sc REFERENCE} model}
\label{sec-ref}
The \reference\ model assumes a cosmologically flat, vacuum energy dominated
$\Lambda$CDM universe with cosmological parameters
$[\Omega_m,\Omega_b,\Omega_\Lambda,\sigma_8, n_s, h]$ $=$ $[0.238, 0.0418,$
$0.762, 0.74,$ $0.951, 0.73]$, as determined from the WMAP 3-year data and
consistent\footnote{Our value of $\sigma_8$ is 8 \% lower than the
  best-fit WMAP 7-year data \citep{Jarosik2010}.} with the WMAP 5-year
data. The assumed primordial helium mass fraction is $Y=X_{\rm He} =
0.248$.  We used \textsc{cmbfast} (version 4.1; \citealt{Seljak1996})
to compute the linear power-spectrum
at the starting redshift $z = 127$. Simulations with given box size use identical initial conditions (phases and amplitudes of the Gaussian density field), enabling us to investigate in detail the effects of the imposed physics on the forming galaxies and the intergalactic medium. The simulations are performed with a gravitational softening that is constant in comoving variables down to $z=2.91$, below which we switch to a softening that is constant in proper units. This is done because we expect two-body scattering to be less important at late times, when haloes contain more particles.

\begin{itemize}
\item {\em Gas cooling and photoionization} Radiative cooling and
  heating are implemented as described in
  \cite{Wiersma2009}\footnote{We used their Eq. (3) rather than
    (4) and \textsc{cloudy} version 05.07 rather than 07.02.}. In
  brief, the radiative rates are computed
  element-by-element,  in the presence of
  an imposed ionizing background and the CMB. We use the redshift-dependent ionizing background
  due to galaxies and quasars computed by \citet[][hereafter
  HM01]{HM01}.  Contributions to cooling and heating of eleven elements
  (hydrogen, helium, carbon, nitrogen, oxygen, neon, magnesium,
  silicon, sulphur, calcium, and iron) are tabulated as a function of
  density, temperature and redshift, using the publicly available
  photo-ionization package \textsc{cloudy}, last described by
  \cite{Feta98}, assuming the gas to be optically thin and in
  (photo-)ionization equilibrium. 

 Hydrogen reionization is implemented by switching on the evolving,
  uniform ionizing background at redshift $z=9$. Prior to this redshift
  the cooling rates are computed using the CMB and a
  photo-dissociating background which we obtain by cutting off the
  $z=9$ HM01 spectrum above 1~Ryd, which suppresses H$_2$ formation and cooling at all redshifts (we do not resolve haloes in which Pop.~III stars that form by H$_2$ cooling).
  
\begin{figure}
  \includegraphics[width=84mm]{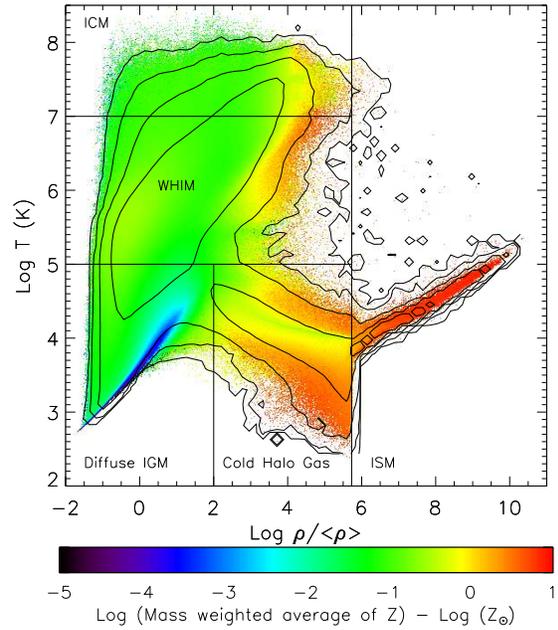}
  \caption{Mass weighted metal distribution in temperature-density space at $z =
  0$ in simulation \textsc{reference\_l100n512}. The colour scale gives the metallicity. The contours indicate the metal mass distribution and are
  logarithmically spaced by 1 dex. The straight lines indicate the adopted division of the gas into: star-forming gas (i.e., $n_{\rm
H}>0.1~\cm^{-3}$), diffuse IGM ($\rho < 10^2\,\langle \rho \rangle$, $T<10^5\,\K$),
  cold halo gas ($\rho>10^2\,\langle \rho \rangle$,
$T<10^5\,\K $), WHIM ($10^5\,\K < T <
10^7\,\K$), and ICM ($T > 10^7\,\K$). The metals are
  distributed over a wide range of densities and temperatures.}
  \label{fig-smZcont}
\end{figure}

\item {\em Star formation} 
The interstellar medium of the Milky Way consists of multiple \lq phases\rq: a warm component in which hot super
nova bubbles envelope and penetrate cold \lq clouds\rq.  Current cosmological galaxy formation simulations cannot resolve
such a multi-phase ISM, and in addition not all the physics that governs the interaction between these phases, and the star formation in them, is included.  Instead of trying to simulate these physical processes, we use the following `sub-grid' model. ISM gas is assumed to follow a pressure-density relation
\begin{equation}
P = P_0(n_{\rm H}/n_{\rm H; thres})^{\gamma_{\rm eff}}\,
\label{eq:EOS}
\end{equation}
where $P_0/k = 1.08 \times 10^3~\cm^{-3}\,\K$. We use $\gamma_{\rm eff} = 4/3$ for which both the Jeans
mass and the ratio of the Jeans length to the SPH kernel are
independent of the density, thus preventing spurious fragmentation due
to a lack of numerical resolution. Finally, only gas dense enough to be
gravo-thermally unstable, $n_{\rm H} \geq n_{\rm H; thres} = 10^{-1}~\cm^{-3}$,
is assumed to be multiphase and star-forming gas \citep{Schaye2004}. It is on this gas that the pressure-density relation
(Eq. \ref{eq:EOS}) is imposed.

Star formation in disk galaxies is observed to follow a power-law `Kennicutt-Schmidt' \cite[]{Kennicutt1998} relation
between the surface density of star formation, $\dot\Sigma_\ast$, and the gas surface density, $\Sigma_{\rm g}$, 
\begin{equation}
\dot{\Sigma}_\ast = 1.5 \times 10^{-4}\, \Msun \yr^{-1} \kpc^{-2} \left
    ({\Sigma_{\rm g} 
    \over 1 ~\Msun \pc^{-2}}\right )^{1.4}\,.
\label{eq:KS1}
\end{equation}
We have implemented this surface density law as a pressure law. A star formation rate which depends on ISM pressure guarantees that simulated disk galaxies in which the disk is vertically in approximate hydrostatic equilibrium follow the observed law, independent of the value of $\gamma_{\rm eff}$ imposed on their ISM gas \citep{Schaye2008}.

\item {\em Galactic winds} Galactic winds are implemented as described
  in \cite{Dallavecchia2008}. Briefly, after a short delay of $t_{\rm
    SN} = 3\times 10^7~\yr$, corresponding to the maximum lifetime of
  stars that end their lives as core-collapse SNe, newly-formed
  star particles inject kinetic energy into their surroundings by
  kicking a fraction of their neighbouring gas particles in a random
  direction, as governed by,
  \begin{eqnarray}
    \dot M_w &=& \eta \dot M_\ast\nonumber\\
    \epsilon_{\rm SN}\,f_w &=& {1\over 2}\,\eta\,v_w^2\,.
    \label{eq:winds}
  \end{eqnarray}
  The mass loading factor, $\eta$, relates the rate at which mass is
  launched in a wind, $\dot M_w$, to the star formation rate, $\dot
  M_\ast$.  The product $\eta\,v_w^2$, where $\eta$ is the mass loading factor and $v_w$ is the initial wind velocity, is proportional to the fraction
  $f_w$ of the SN energy produced per unit mass, $\epsilon_{\rm SN}$,
  that powers the wind. We usually characterise the wind implementation
  by the mass loading factor, and wind speed. 

The wind prescription is
  implemented as follows: each SPH neighbour $i$ of a newly-formed star
  particle $j$ has a probability of $\eta m_j/\sum_{i=1}^{N_{\rm
      ngb}}m_i$ of receiving a kick with a velocity $v_{\rm w}$. Here,
  the sum is over the $N_{\rm ngb} = 48$ SPH neighbours of particle $j$. The
  \reference\ simulation described below has a mass loading of $\eta =
  2$ and wind speed of $v_{\rm w} = 600~\kms$ (i.e., if all baryonic
  particles had equal mass, each newly formed star particle would kick,
  on average, two of its neighbours, increasing their velocity by
  $v_{\rm w} = 600~\kms$). Assuming that each star with initial mass in
  the range $6-100~\Msun$ injects $10^{51}~\erg$ of kinetic energy as
  it undergoes a core collapse SN, these parameters imply that
  the total wind energy accounts for 40 per cent of the available
  kinetic energy for a Chabrier IMF and a stellar mass range
  $0.1-100~\Msun$ (if we consider only stars in the mass range
  $8-100~\Msun$ for type II SNe, this works out to be 60 per cent). The
  value $\eta=2$ was chosen to roughly reproduce the peak in the cosmic
  star formation rate \citep{Schaye2010}. Note that contrary to the widely-used kinetic
  feedback recipe of \cite{Springel2003}, the kinetic energy is
  injected \emph{locally} and the wind particles are \emph{not}
  decoupled hydrodynamically. As discussed by \cite{Dallavecchia2008},
  these differences have important consequences.

\begin{table*}
\caption{Simulation Set}
\label{tab-simset}
\begin{tabular}{lrll}
\hline
Name & \multicolumn{2}{c}{Box Size (Mpc/h)} &  Description  \\ \hline
AGN & 25 & 100 & Incorporates AGN model of \cite{Booth2009}.\\
DBLIMFCONTSFV1618 & 25 & 100 & Top-heavy IMF for $n_{\rm H} > 30~{\rm  cm}^{-3}$; $v_{\rm w} = 1618~{\rm km}\, {\rm s}^{-1}$\\
DBLIMFV1618 & 25 & 100 & Top-heavy IMF for $n_{\rm H} > 30~{\rm cm}^{-3}$; $v_{\rm w} = 1618~{\rm km}\, {\rm s}^{-1}$, $\dot\Sigma_\ast(0)=2.083\times 10^{-5}\Msun \yr^{-1} \kpc^{-2}$\\
DBLIMFCONTSFML14 & 25 & 100 & Top-heavy IMF for $n_{\rm H} > 30~{\rm cm}^{-3}$; $\eta = 14.545$ \\
DBLIMFML14 & 25 & 100 & Top-heavy IMF for $n_{\rm H} > 30~{\rm cm}^{-3}$; $\eta = 14.545$, $\dot\Sigma_\ast(0)=2.083\times 10^{-5}\Msun \yr^{-1} \kpc^{-2}$\\
EOS1p0 & 25 & 100 & Isothermal equation of state \\
EOS1p67 & 25 && Equation of state $p \propto \rho^{\gamma_{\ast}}$, $\gamma_{\ast} = 5/3$\\
IMFSALP & 25 & 100 & Salpeter IMF, SF law rescaled \\
MILL & 25 & 100 & WMAP1 cosmology: $(\Omega_m,\Omega_{\Lambda},\Omega_b h^2,h,\sigma_8,n,X_{\rm He}) = (0.25,0.75,0.024,0.73,0.9,1.0,0.249) $ \\
NOAGB\_NOSNIa && 100 & AGB \& SNIa mass \& energy transfer off \\ 
NOHeHEAT & 25 && No He reheating \\
NOSN & 25 & 100 & No SNII winds, no SNIa energy transfer \\
NOSN\_NOZCOOL & 25 & 100 & No SNII winds, no SNIa energy transfer, cooling uses primordial abundances \\
NOREION & 25 && No reionization \\
NOZCOOL & 25 & 100 & Cooling uses primordial abundances \\
REFERENCE & 25 & 100 & \\
REIONZ06 & 25 && Redshift reionization = 6  \\
REIONZ12 & 25 && Redshift reionization = 12  \\
SFAMPLx3 & 25 && $\dot\Sigma_\ast(0) = 4.545\times 10^{-4}\Msun \yr^{-1} \kpc^{-2}$ \\
SFSLOPE1p75 & 25 && $\gamma_{\rm KS} = 1.75$ \\
SFTHRESZ & 25 && Metallicity-dependent SF threshold \\
SNIaGAUSS && 100 & Gaussian SNIa delay distribution  \\
WDENS & 25 & 100 & Wind mass loading and velocity determined by the local density \\
WHYDRODEC & 25 &&  Wind particles temporarily hydrodynamically decoupled \\
WML1V848 & 25 & 100 & $\eta = 1$, $v_{\rm w} = 848 \kms$  \\
WML4 & 25 & 100 & $\eta = 4$, $v_{\rm w} = 600 \kms$  \\
WML4V424 & 25 && $\eta = 4$, $v_{\rm w} = 424 \kms$  \\
WML8V300 & 25 && $\eta = 8$, $v_{\rm w} = 300 \kms$ \\
WTHERMAL & 25 && SNII energy injected thermally \\
WVCIRC & 25 & 100 & `Momentum driven' wind model (scaled with the resident halo mass)\\
\hline
\end{tabular}
\end{table*}

\item {\em Chemodynamics} We employ the method outlined in \cite{Wiersma2009b}. 
  Briefly, a star particle forms with the elemental abundance of its
  parent gas particle. It then represents a single stellar population (SSP)
  with given abundance, and an assumed stellar IMF. 
  The reference model uses the IMF proposed by \cite{C03}, with mass limits of $0.1 \Msun$ and $100 \Msun$. At each time step, we
  compute the timed release of elements and energy from three stellar evolution channels: (i) core collapse SNe, (ii)  type I SNe and (iii) asymptotic giant branch (AGB) stars. The
  stellar evolution prescriptions are based on the Padova models, using
  stellar lifetimes computed in \cite{Peta98} and the yields of low-mass and high-mass stars of \cite{M01} and \cite{Peta98} respectively. The yields
  of \cite{Peta98} include ejecta from core collapse SNe (SNII)
  along with their calculations of mass loss from high-mass
  stars. Using these yields gives an element of consistency between the
  high- and low-mass stellar evolution.  An e-folding delay time is used to describe the SNIa rate. The observed cosmic SNIa rate is approximately reproduced in the \reference\ model with a fraction of $2.5\%$ of white dwarfs that become SNIa, see \cite{Wiersma2009b}.  We use the `W7' SNIa yields of \cite{Teta03}.

  Elements produced by nucleo-synthesis are distributed to SPH
  particles neighbouring the star, weighted by the SPH kernel, as done
  by e.g., \cite{Mosconi2001}.  The simulations track the
  abundance of the eleven individual elements that are important for the cooling and we use an extra \lq metallicity\rq\
  variable to track the total metal mass of each particle \cite[see][]{Wiersma2009b}. The ratio $Z\equiv M_Z/M$ of metal mass over total mass is
  the \lq particle metallicity\rq\ .  For gas particles we also compute
  a \lq smoothed metallicity\rq\, as $Z_{\rm sm} \equiv \rho_{\rm Z}/ \rho$, i.e.\ the ratio of the metal density, $\rho_{\rm Z}$, over the
gas density, $\rho$, where both densities are computed using SPH interpolation. Stars inherit metal abundances of their parent gas
particle: we record both the particle and smoothed metallicity. In \cite{Wiersma2009b} we
argued that the smoothed metallicity is more
consistent with the SPH formalism than the particle metallicity. Using smoothed metallicities results 
in a spreading of metals over slightly greater volumes. 
\end{itemize}

\begin{figure*}
  \scalebox{0.85}{\includegraphics[width=76mm]{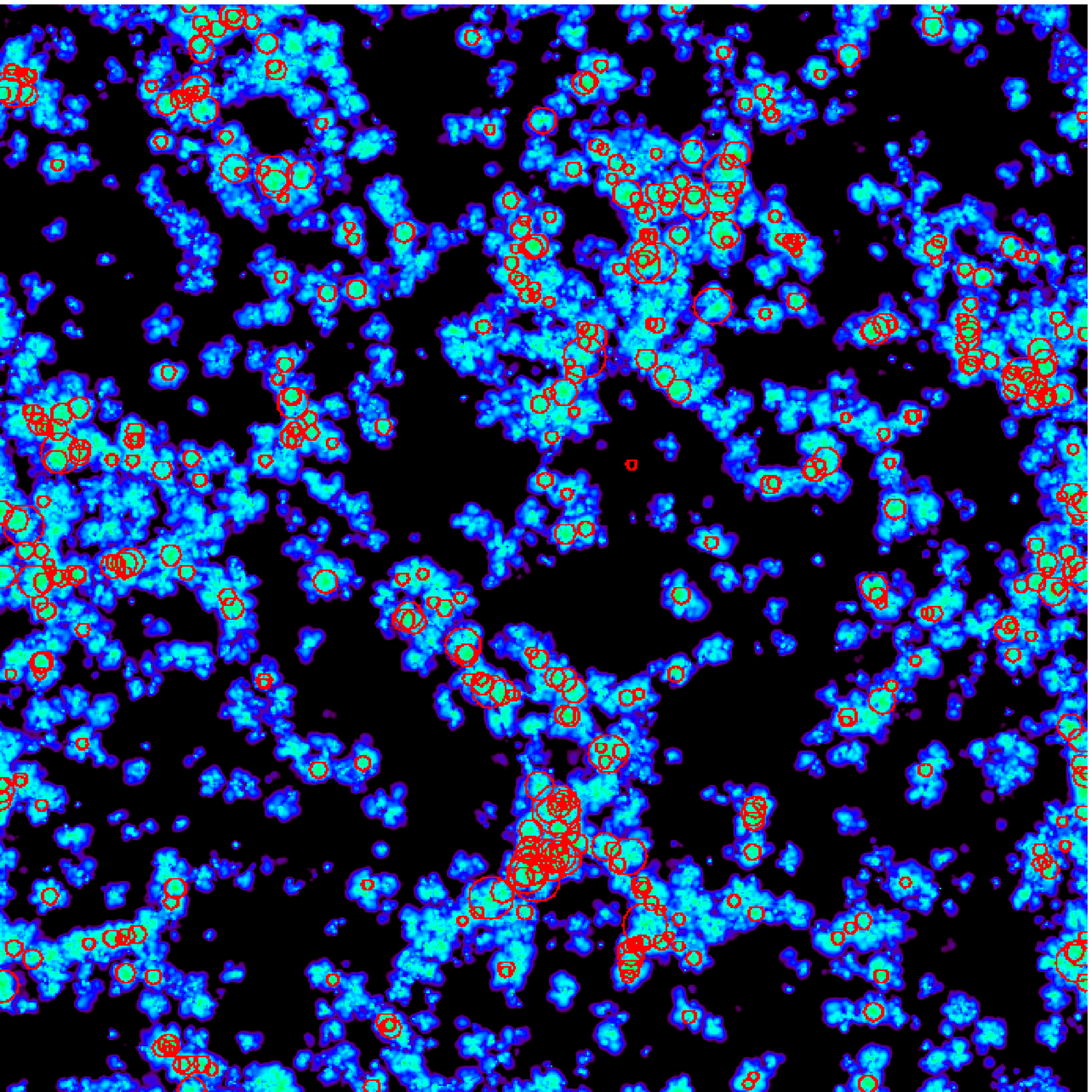}
  \includegraphics[width=76mm]{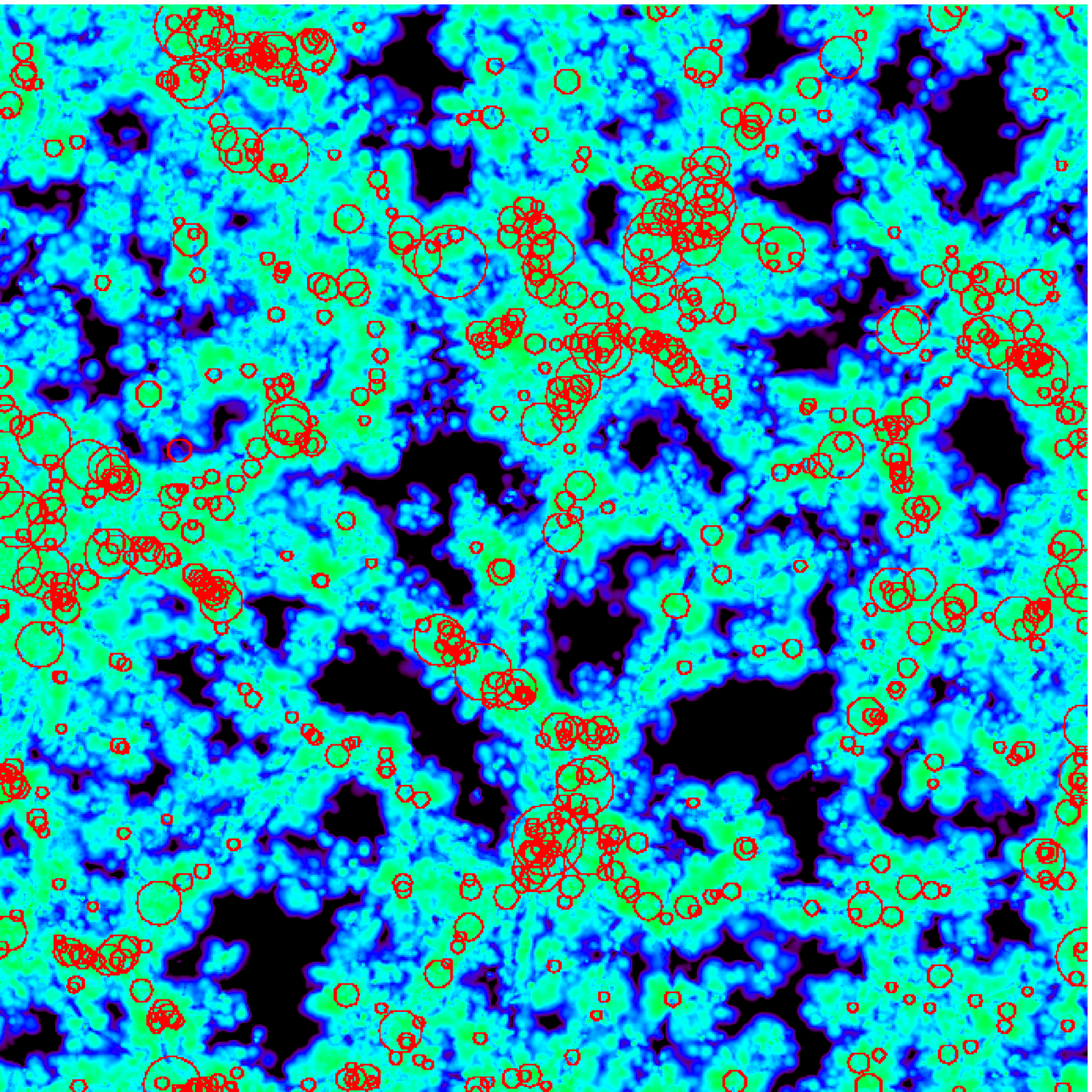}
  \includegraphics[width=15mm]{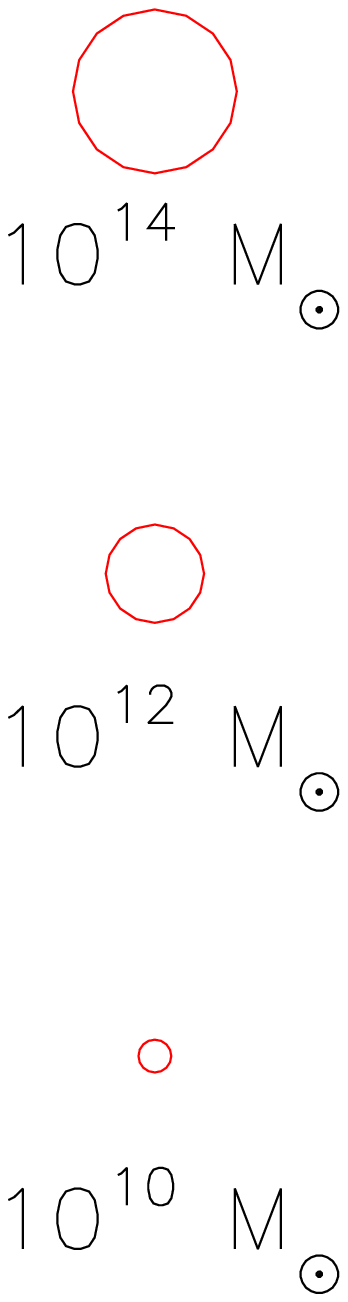}}\\
\vspace{0.07cm}
  \scalebox{0.6}{\includegraphics[width=168mm]{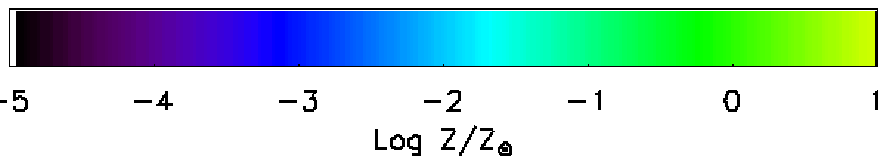}}
\caption{Metal distribution in the reference simulations. 
  Shown are
$5 \hMpc$ (comoving)
  thick slices through the \textsc{reference\_l025n512} simulation (top row) at $z = 4$
  (left), and $z = 2$ (right) and through the \textsc{reference\_l100n512} simulation
(bottom row) at $z = 2$
  (left), and $z = 0$ (right).
  The colour coding shows (SPH-smoothed) metallicity, mass averaged along the line of sight; the colour scale is cut off below $\log(Z/Z_{\odot})$ = -5 although metallicities extend to lower values.  Red circles correspond to haloes identified by a friends-of-friends group finder, with radius proportional to the logarithm of the stellar mass of the
  halo, as indicated to the right of the panels. Metals are initially strongly clustered around haloes, but their volume filling factor increases as time progresses.}
\label{fig-pretty1}
\end{figure*}

\subsection{The Simulation Suite}
Table~\ref{tab-simset} contains an overview the models from the \owls\ suite that we consider here. 
Simulation names contain a string \lq LXXXNYYY\rq\ which specifies the co-moving  size of the periodic box, L=XXX$h^{-1}$~Mpc, and the number of particles N=YYY$^3$ of (initial) gas and dark matter; most runs discussed here have L=25 or 100, and N=512$^3$. Comparing different L and N models allows us to investigate the effects of numerical resolution, and missing large-scale power. The table also contains a brief description of the physics in which a particular run differs from the \reference\ model, see \cite{Schaye2010} and the next section for more details.  

\begin{figure*}
\includegraphics[width=168mm]{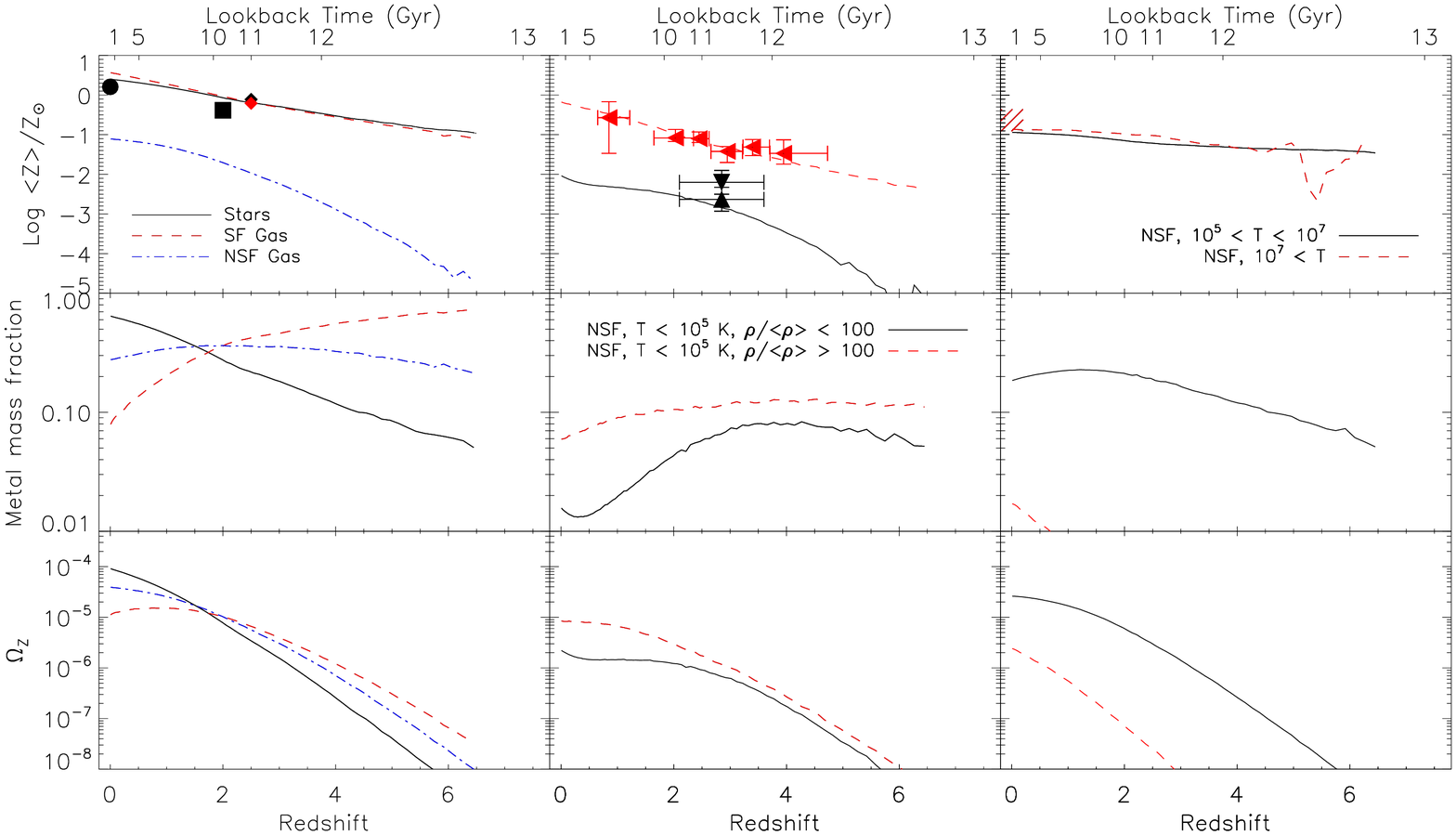}
\caption{Metallicity (top row), metal mass fraction (middle row), and
  $\Omega_Z$ (bottom row) as a function of redshift for the various
  phases of baryons in the \textsc{reference\_l100n512} simulation. Curves indicate simulation results, with different line styles indicating various baryonic phases. {\em Left column}:  stars (solid black), SF gas (dashed red), NSF gas (dot-dashed blue); {\em Middle column}: diffuse IGM (black), cold halo gas (dashed red); {\em Right column}: WHIM (solid black line), ICM (dashed red).
  Data points indicate observations and are colour-coded to indicate to which phase they should be compared.  The diamonds at 
  $z = 2.5$ are estimates of stellar and star-forming metallicities by \protect\cite{Pagel2008}, with further stellar metallicities indicated by a circle \protect\citep{Gallazzi2008} and a square
  \protect\citep{Halliday2008}. DLA measurements
  are from \protect\citet[solid, left pointing triangles]{Prochaska2003}, IGM pixel optical depth measurements in QSO spectra are from \protect\citet[open triangle pointed down]{Aguirre2008} and
  \protect\citet[open triangle pointed up]{Seta03}. The ICM measurements
  are from X-ray observations (\protect\citealp[hatched
    region]{Simionescu2009}).}
     \label{fig-killer}
 \end{figure*}

\section{Results}
\label{sec-WATM}

\subsection{The reference model}

We begin by describing the metal distribution of the
\textsc{reference\_l100n512} simulation in temperature-density space at
redshift $z=0$ shown in Fig.~\ref{fig-smZcont}. 

It is useful to divide the $T-\rho$ plane into several `phases', since,  as we will show, different enrichment processes dominate in different phases, and also because the observational constraints for different phases are inferred from different types of data.
A first division distinguishes between star-forming gas (henceforth SF gas) on
the imposed $P-\rho$ relation (Eq. \ref{eq:EOS}), which we identify
with the ISM in galaxies, and non-star-forming gas (NSF gas). The SF
gas can be seen as the thin band of contours on the right of
figure \ref{fig-smZcont} at $\rho/\langle\rho\rangle>10^6$. We further
divide NSF gas into {\em hot gas}, typically found in halos of large
groups or clusters (ICM, $T > 10^7\,\K$), {\em warm-hot intergalactic} and {\em
  circum-galactic gas} (WHIM, $10^5\,\K <T < 10^7\,\K $), 
{\em diffuse gas} (diffuse IGM, $\rho < 10^2 \langle \rho
\rangle$, $T < 10^5\,\K $), and {\em
  cold halo gas} ($\rho>10^2 \langle \rho \rangle$, $T< 10^5\,\K$). These phases are indicated 
in Fig.~\ref{fig-smZcont}. The colour coding indicates the mass-weighted 
mean metallicity for a given temperature and density. The metallicity shows a 
clear gradient with density, with the exception of the track of photo-ionised gas 
in the lower left hand corner, which has a very low metallicity. The contours, 
on the other hand, show the distribution of total metal mass. Metals are 
found at virtually
all temperatures and densities, from the high-densities in galactic
disks, to extremely low densities.

The spatial distribution of metals is shown at $z=4$, 2 and 0, in
Fig.~\ref{fig-pretty1}. At $z=4$ (top left) metals are strongly clustered around haloes, with circum-halo metallicities of $\log(Z/Z_\odot)\approx -3$ to -2,
and large fractions of volume are enriched to exceedingly low levels, or not at all.
As time progresses, circum-halo metallicities increase to $\log(Z/Z_\odot)\approx -2$ to -1 and the filling factor of metals also increases substantially, yet even at $z=0$ (bottom right) there are still co-moving volumes which are barely enriched.
Comparing the two resolutions (and box sizes), we note that although the abundance of circum-halo gas is similar at $z=2$ in the L100N512 (bottom left) and L025N512 (top right) runs, the filling factor of metals is higher in the higher resolution simulation. Indeed, we showed in \citet{Wiersma2010} that lower density gas (which accounts for larger filling factors) was typically enriched by lower-mass galaxies, and higher resolution simulations probe the galaxy mass function down to lower masses.

We continue our analysis by comparing the metallicities and metal mass fractions in different phases provided by the reference simulation with observations. We then turn our attention to a comparison of the different \owls\ models.

\begin{figure*}
\scalebox{0.85}{\includegraphics[width=84mm]{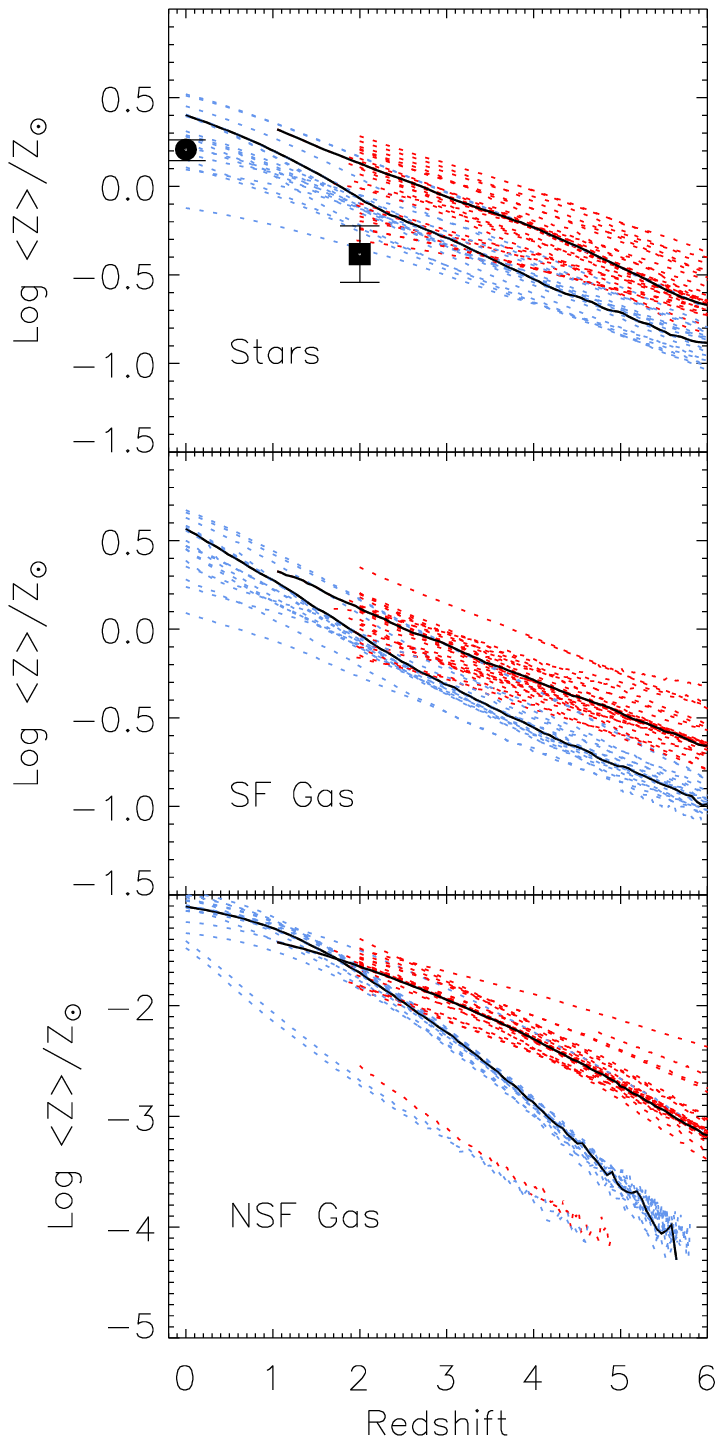}\hspace{1cm}
  \includegraphics[width=84mm]{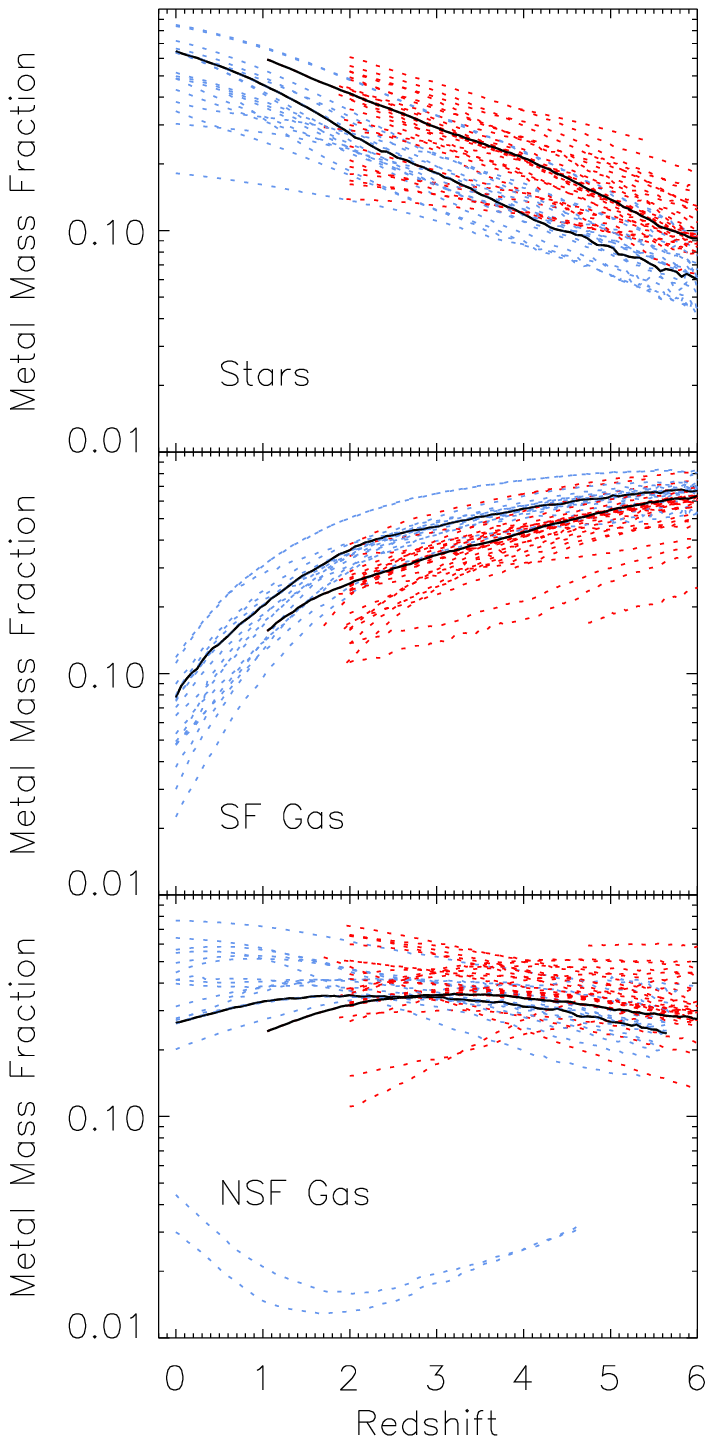}}
  \caption{Mean metallicity (left) and metal mass fraction (right) for stars (top), star-forming gas (middle),
        and non-star-forming gas (bottom) as a function of redshift for
        all \owls\ simulations. Shown are the \reference\ simulations
        (solid black), the rest of the \textsc{l025n512} simulations 
        (dotted red,) and the rest of the \textsc{l100n512} simulations (dotted blue). The higher resolution (\textsc{l025n512}) simulations
        were stopped at $z=2$ (with the exception of the \reference\ model which was stopped just above $z  = 1$), but the lower resolution simulations
        (\textsc{l100n512}) were continued to $z=0$.  The mean stellar
        metallicity in units of solar rises from $\sim -0.5$ dex at
        $z=6$ to supersolar at $z=0$, closely tracking the
        metallicity of the SF gas. For these two phases, resolution
        plays a minor role, with a 64 times higher mass resolution resulting in about 0.2 dex higher
        metallicities. The metallicity of NSF gas
        is, however, strongly resolution dependent for $z>2$. In the higher
        resolution simulation it rises from Log$\langle Z\rangle
        /Z_{\odot} \approx -3$ at $z=6$ to to $-1$ at $z=0$.  Points with
        error bars indicate observations of stellar metallicities at
        $z = 0$ and $z = 2$ by \protect\cite{Gallazzi2008} and
        \protect\cite{Halliday2008}, respectively. The
        stellar metal mass fraction increases towards $z=0$, whereas
the metal
        mass fraction in SF-gas decreases slowly to $z\approx 2$, then
        falls much more rapidly. Aside from the
        models without feedback (the group of very low curves in the bottom panels; note that \textsc{nosn\_l025n512} and \textsc{nosn\_nozcool\_l025n512} fall on top of each other in the left panel and fall below the plotting area in the right panel),
        most of the different runs yield results that follow
        the \reference\ model relatively closely, indicating that the
        predicted metallicities are reasonably robust to
        model variations.}
      \label{fig-allevol}
\end{figure*}

Fig.~\ref{fig-killer} shows the evolution of the metallicity\footnote{Note that for the remainder of this paper, we use
`particle metallicities' (see section \ref{sec-ref}), although since we
are considering mean metallicities of a given phase, the results would have been
virtually identical if we had used smoothed metallicities.}, metal mass fraction, and $\Omega_Z$ for various phases in the \textsc{reference\_l100n512} simulation. The fraction of metals in a given phase is simply the ratio of the
metal mass in that phase over the total metal mass in the simulation. The amount of metals in a phase is shown as $\Omega_Z
\equiv \left < \rho_Z\right >/\rho_{\rm crit}$, the ratio of the mean metal mass density in
a given phase over the critical density. As the metal mass fraction and $\Omega_Z$ are much harder to measure observationally, and are in any case inferred from metallicity measurements, we only compare the predicted metallicities with observations and will mostly focus on metallicities in the rest of this work.

Now we briefly walk through the data (which we have converted to our adopted 
solar abundances) to which we are comparing:
\begin{itemize}
\item {\em Stars} For stellar metallicities, we compare to the global
  values obtained by \citet[circle in the top-left panel]{Gallazzi2008} for $z = 0$ and \citet[square in top left panel]{Halliday2008} for $z = 2$. \citet{Halliday2008} note that their metallicities
  are lower than those of \citet{Eeta06} and speculate that
  since \citet{Eeta06} observed oxygen, a non-solar [O/Fe] ratio might make
  their observations agree.  We have also compared their point to
  [Fe/H] and indeed find a better match to the data. The $z = 2.5$ point is taken from \citet{Bouche2007}
  which is compiled from their own previous measurements. 
\item {\em Star-forming gas} \citet{Bouche2007} give a value for ISM and/or dust at $z = 2.5$ and, 
  following \citet{Pagel2008}, we include it here as a filled red diamond in the top-left panel.
\item {\em Diffuse IGM} We have included
  the carbon and oxygen Lyman-$\alpha$ forest measurements from
  \citet[upwards pointed triangle in the top middle panel]{Seta03} and \citet[downwards pointed
    triangle in the top middle panel]{Aguirre2008},
  respectively. 
\item {\em Cold Halo Gas} While the precise nature of DLAs is still
  somewhat uncertain, we surmise that circum-halo and intrahalo cold
  gas most likely have higher cross-sections than the ISM
  and we thus assume that observations of DLA and sub-DLA systems trace cold halo gas. As such, we have plotted the
  compilation of DLA metallicities from \citet[leftwards pointed
    triangles in the top middle panel]{Prochaska2003}.
\item {\em WHIM} The WHIM has long stood as a phase that is thought to contain a large fraction of the baryons and metals, but is difficult to 
  detect. As such, there is no direct measurement of the metallicity of the WHIM. 
\item {\em ICM} Measurements of ICM metallicities are also troublesome,
  although much less so than for the WHIM. The main difficulty is measuring the
  metallicity out to large enough radii in order to get a good
  estimate of the mean metallicity. We have chosen to use the
  outermost measurements of cluster metallicities by \citet[hatched
    region in the top right panel]{Simionescu2009}, using a box to
  indicate the range of values in their sample. 
\end{itemize}

\begin{figure*}
\includegraphics[width=168mm]{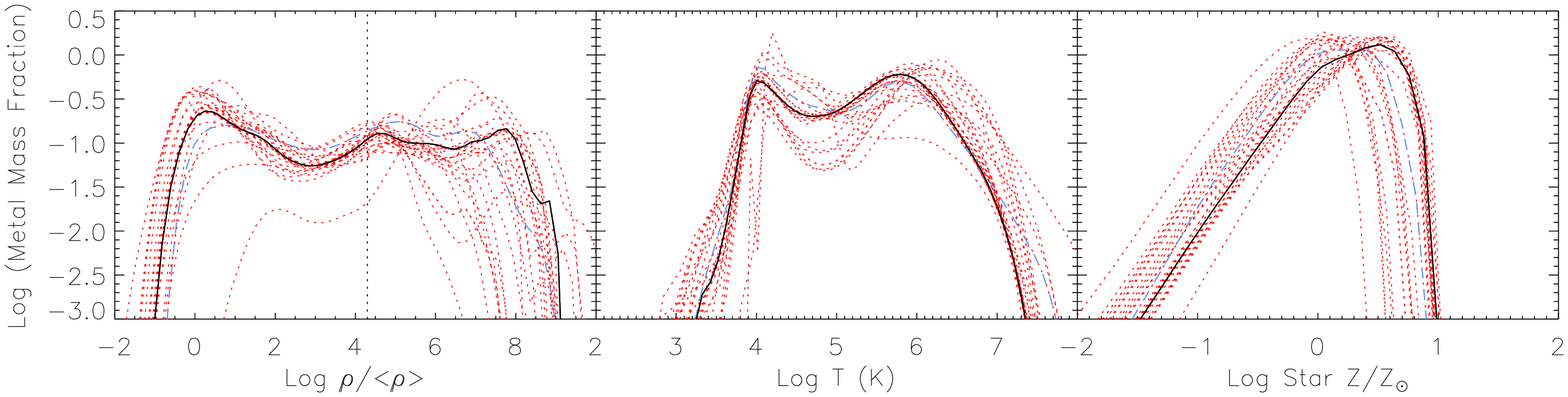}\\ 
\includegraphics[width=168mm]{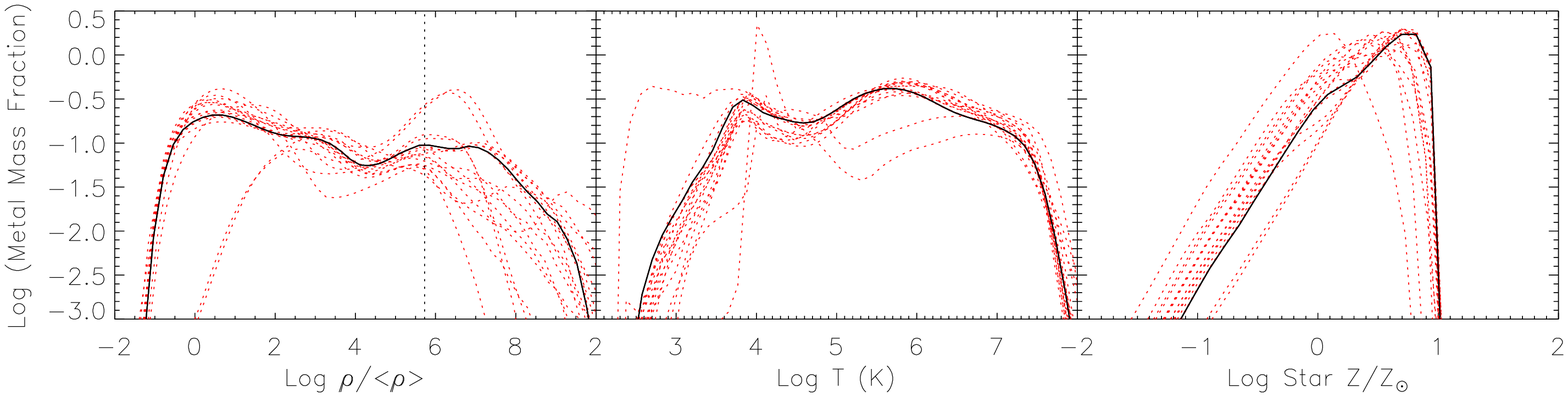}
\caption{The left and middle panels show the fraction of metals in gas as a function of density and temperature, respectively, and the right panels shows the fraction of metals in 
  stars as a function of stellar metallicity. The top and bottom rows show $z=2$ and 0, respectively. Each histogram is normalised such that it sums to unity. The reference 
simulations are shown as solid black, while the rest of the simulations are 
shown in dotted red. The \textsc{reference\_l100n512} model at $z = 2$ is 
shown in
  the top panel with the dashed blue line. SF gas is omitted
  from the central panel, since the `temperature' of this
  multi-phase gas is set by the imposed $P-\rho$ relation.  The dotted
  vertical lines indicate the star formation threshold. While there are a few 
  outliers, most models yield similar metal distributions, with the largest differences occurring at high $\rho$ and small $T$. }
\label{fig-all1D}
\end{figure*}

The metallicities of our simulation follow the trends seen in the data. We reproduce the 
observed increase with cosmic time of the stellar and cold halo metallicities (although our stellar metallicities 
fall slightly above the observed points). The high-redshift measurements for the star-forming 
gas are satisfied as well. The predicted metallicity of the $z\approx 3$ diffuse IGM agrees with the measurement for carbon \citep{Seta03}, but is about 0.5 dex lower than the observed value for oxygen \citep{Aguirre2008}, which should be closer to the overall metallicity. 
Our prediction for the intracluster metallicity is within the observed range.

\subsection{Overview}
We want to investigate which physical processes determine the cosmic metal distribution in a given phase. We begin
by discussing some of the overall trends that are present in nearly all of our models. Fig.~\ref{fig-allevol} shows the evolution of the metallicity (left) and the metal mass fraction (right) for stars (top), SF gas (middle), and NSF gas (bottom). The black curves show the results for the reference model, with the curve that continues down to $z=0$ indicating the 100 Mpc/h box. Blue and red curves indicate the results for the 100 and 25 Mpc/h boxes for all other models.

The metallicities of the SF gas and stars track each other very
closely, increasing steadily from $\log(Z/Z_\odot)\sim -0.5$ at $z=6$
to supersolar values at $z=0$. While the metal mass fraction contained
in stars increases with time, from $\sim 10^{-1}$ at $z=6$ to 0.2-0.9
at $z=0$, the fraction of the metal mass contained in SF gas decreases
with time, and does so more rapidly for $z<2$. All models predict that
by $z=0$ SF gas accounts for only a small fraction of the metal mass.
The rapid decrease in the metal mass fraction in SF gas for $z<2$
reflects the decrease in the cosmic star formation rate
\citep{Schaye2010} and hence the global mass density of SF gas. While
the metallicity of the NSF increases rapidly with time, from $\la
10^{-3}\,Z_\odot$ at $z=6$ to $10^{-1}\,Z_\odot$ at $z=0$, the
fraction of the metals residing in this diffuse phase is relatively
constant and, except for models without any feedback, always
significant (for both reference simulations this fraction stays between 20\% and 35\% between $z=6$ and $z=1$). Initially, the majority of the metals reside in SF
gas. At $z\approx 3$ metals are approximately equally distributed over
the three different phases (stars, SF and NSF gas), with stars
becoming the dominant repository of metals at lower redshift.

Comparing the two different box sizes (blue vs red curves), for which the mass resolution differs by nearly two orders of magnitude, we see that the results for stars and SF gas are surprisingly (given the enormous change in resolution) well converged. The higher resolution runs predict stellar metallicities and metal mass fractions that are about 0.2 dex higher. For SF gas the metallicity is also slightly higher (about a factor of two at $z=6$ and less at lower redshifts), but the metal mass fraction is slightly lower, because there is more gas in this phase in higher resolution runs. The metallicity of NSF gas is much more sensitive to the resolution and is
unreliable in the lower resolution simulations above $z\approx 2$. The fact that the two different resolutions yield similar metal mass fractions for NSF gas at high redshift is a coincidence, caused by the fact that there are far fewer metals in the lower resolution runs. 

We recall the results of the more extensive numerical convergence tests performed for the reference model in appendices B
and C of \cite{Wiersma2009b}. For metallicity and metal mass fraction, except possibly for the ICM, our simulation boxes are sufficiently large. On the other hand, resolution proves to be
much more of a challenge. The stellar metal mass fraction just barely
converges for the \textsc{l025n512} resolution, although the difference is small by $z = 2$. The metal mass fraction in the cold-phase NSF gas is converged to within a factor of two. Obtaining converged results tends to be more challenging at higher redshifts. Metallicity is generally better converged than the fraction of metals in a given phase, and is reasonably reliable for all phases at
the resolution of the \textsc{l100n512} simulations, with the exception of the metallicity of the diffuse IGM. Higher resolution simulations generally yield higher metallicities, especially at higher redshift, but even in this phase, the simulations are converged by $z = 2$ ($z = 3$) for \textsc{l100n512} (\textsc{l025n512}) runs.

The distributions of metals at $z=0$ and $z = 2$ are investigated in
more detail in Fig.~\ref{fig-all1D}. The metal mass weighted probability
distribution function (PDF) \footnote{We emphasise that the right most panel does {\it not} 
show the metallicity distribution function, as is commonly plotted. It rather shows 
how the metal mass is distributed among stars of various metallicities, which is more heavily weighted towards higher metallicities than the metallicity distribution function.} shows several maxima in both density and
temperature, with minima at $\rho \sim 10^3 \langle\rho\rangle$ and
 $\rho \sim 10^4$ at $z=2$ and $z=0$ respectively, which corresponds to $n_{\rm H} \sim 10^{-3}\,\cm^{-3}$ in both cases, and at $T \approx 10^5\,\K$. 

All models have a relatively well pronounced maximum at
$\rho \sim \langle\rho\rangle$, which is due to the fact that there are very few metals in underdense gas and that gas that is much denser quickly accretes onto haloes.
The outliers at the low-density end of the distribution 
result from turning off feedback (see Fig.~\ref{fig-zcool0metevol}), while the outliers at the high-density end correspond to either 
strong feedback models (e.g.\ \agn) or models in which the star formation efficiency is increased (e.g.\ \textsc{sfampl{\small x}3}).

The temperature minimum in the metal mass fraction at $T \sim 10^5\,\K$ is caused by a number of factors. First, gas cooling is very
efficient at this temperature. Second, the winds tend to shock metals
to temperatures higher than this value, causing a peak in the
distribution at higher temperatures. Finally, the equilibrium
temperature due to photo-heating is lower than this temperature,
causing a second peak at $T\sim 10^4\,\K$. The temperature distribution of the metals
shows less variation than the distribution with density. The two outliers
again correspond to models without feedback (see Fig.~\ref{fig-zcool0metevol}). Such models have little shock-heated metals,
resulting in much more low temperature metals. Without metal cooling,
these metals pile up at $10^4~\K$, whereas with metal cooling the
metals are found to very low temperatures. This effect is diminished
at $z = 2$, although note that the \textsc{nosn\_l025n512} simulation
is not shown because it was stopped at $z \approx 3$ due to computational costs. 

The distribution of metals over stars with different metallicities (right panels) is similar in the different models, with the main deviation at high
metallicity being due to the \AGN\ simulation, which predicts more metal mass in lower metallicity stars.

These figures once more illustrate the large
dynamic range in density and temperature over which metals are
distributed. These distributions are mostly converged with respect to both box size and resolution \cite[Figs.~B3 and C3 of][]{Wiersma2009b}, with some dependence of the temperature distribution on box size, and of the density distribution on resolution.
Next, we will investigate in more detail how various physical processes affect the metal distribution.

\subsection{Impact of galactic winds and metal-line cooling}
\label{sec-feed}

\begin{figure*}
  \includegraphics[width=168mm]{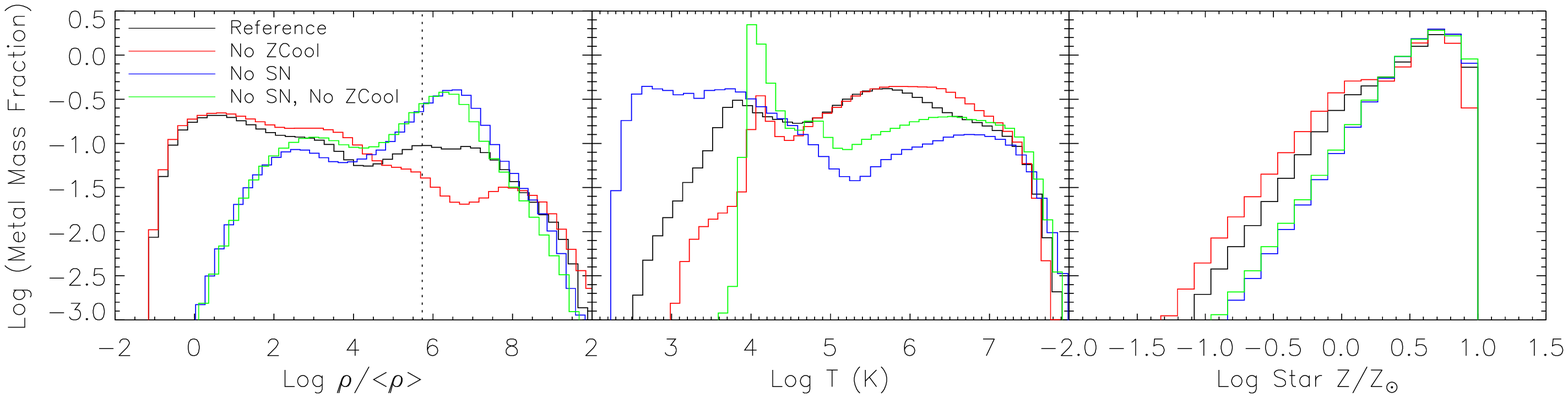}\\ 
  \includegraphics[width=168mm]{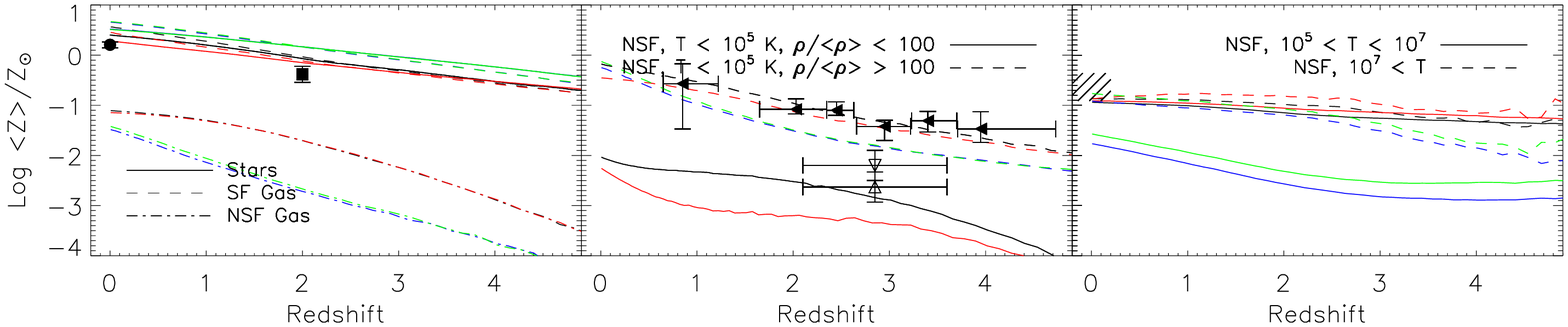}
  \caption{Distribution of metals for {\sc l100n512} simulations with 
    different feedback and cooling
    prescriptions. {\em Top panels}: $z = 0$ metal mass weighted PDFs of 
    gas density, temperature, and stellar metallicity. Each PDF is normalised to unity. In the left-most panel,
    the star formation threshold ($n_{\rm H}=10^{-1}\,\cm^{-3}$) is indicated by the dotted line. ISM gas was removed from the temperature PDF.
    {\em Bottom panels} : Metallicity as a function of redshift for the 
    phases as defined in Fig.~1. The left panel shows stars (solid), SF gas
    (dashed) and NSF gas (dot-dashed). The middle panel shows cold IGM 
    (NSF, $\rho < 10^2 \langle \rho \rangle$, $T < 10^5\,\K$; solid), and 
    cold halo and
    circum-halo gas (NSF, $\rho > 10^2 \langle \rho \rangle$, $T <
    10^5\,\K$; dashed). Finally, the right panel shows the WHIM (NSF gas, 
    $10^5\,\K < T < 10^7\,\K$; solid)
    and the ICM (non-star-forming, $T > 10^7\,\K$; dashed). Shown are the
    \reference\ simulation (black), a simulation which ignored metal cooling ({\sc nozcool}; red), a simulation without galactic winds driven by SNe (\NOSN; blue), and a model with neither winds nor metal cooling ({\sc nosn\_nozcool}; green), as indicated in the top left panel. Note that the metallicity of the diffuse IGM for the simulations without feedback is 
    below the plotted range in the bottom-middle panel.
    Data points indicate observations as in Fig.~\protect\ref{fig-killer}, where the points in the left panel correspond to stellar metallicities, 
    the solid triangles in the middle panel correspond to the cold 
    halo gas while the open triangles show diffuse IGM 
    measurements and the hatched region in the right panel shows an 
    ICM metallicity measurement. Metal cooling shifts the metal distribution towards lower temperatures and increases the fraction of metals in the ISM. Galactic winds are crucial for the enrichment of low-density gas. The metallicity of the ICM is insensitive to both metal cooling and galactic winds.}
\label{fig-zcool0metevol}
\end{figure*}

In this section we discuss the effect of including galactic winds
driven by SNe and metal-line cooling on the metal distribution. We show that
outflows are essential for enriching the
IGM and that metal-line cooling strongly affects the temperature distribution of metals. On the other hand, the hot non-star-forming gas can
obtain rather high metallicities without feedback, indicating that the
ICM is enriched at least partly by some process that is not directly
related to SN feedback.

Fig.~\ref{fig-zcool0metevol} compares the 
\reference\ simulation to a simulation
without metal cooling (\NOZCOOL), one without winds (\NOSN; note that the blue lines in 
the lower left panel fall mainly under the green lines), and one
with neither metal cooling nor winds (\NOSNNOZCOOL). These
simulations differ not only in the way metals are distributed,
but also in their star formation histories since cooling and feedback of
course also affect the conversion of gas into stars \citep[see][]{Schaye2010}.

The top panels shows metal mass weighted PDFs at $z = 0$. The simulation 
without metal cooling contains fractionally more metals than the \reference\ 
model at $\rho \sim 10^2\,\langle\rho\rangle - 10^4\,\langle\rho\rangle$ at the expense of metals in the ISM. Metal cooling thus allows
metals to condense from haloes into star-forming gas. 
This clearly illustrates the importance of
metal cooling for gas accretion onto star-forming regions. Surprisingly, 
the presence of metal cooling has little effect on the temperature 
distribution where the cooling curve peaks ($T \approx 10^5~\K$). This is most 
likely because the cooling times here are already short. Metal cooling does reduce the fraction of metals at $T\sim 10^{6.5}\,\K$. The largest difference 
is noticed in the low temperature regions (we remind the reader that the temperature distribution excludes gas with densities above our star-formation threshold). Metals 
can cool gas well below $10^4\,\K$, while neglecting metal cooling abruptly cuts the distribution off, although there are still some metals with $T< 10^4\,\K$ due to adiabatic cooling. Finally, metal cooling slightly increases the fraction of metals 
locked up in high metallicity stars.

In the absence of galactic winds, only a negligible fraction of the
metals reach densities $\rho \la 10~\langle\rho\rangle$ (top-left
panel of Fig.~\ref{fig-zcool0metevol}).  This underlines the
importance of feedback in reproducing the metals seen in the
IGM. Note that this is in opposition to
\cite{Gnedin1998} who found that feedback played only a minor role in
enriching the IGM, but agrees well with \cite{Aguirre2001a, Aguirre2001b}.
Winds also serve to evacuate metals from the ISM as we can see that in 
the absence of feedback, metals tend to pile up just above the 
star-formation threshold. Without winds, the fraction of metals at (\textsc{nosn\_nozcool}) or below (\textsc{nosn}) $10^4\,\K$ is much
higher than in any model that includes feedback 
from SNe (top-middle panel of 
Fig.~\ref{fig-zcool0metevol}), because the metals reside in gas with higher densities and metallicities and thus higher cooling rates. Winds strongly boost the fraction of metals in gas with $10^5\,\K < T < 10^7\,\K$, which suggests that wind shocks may also shape the metal distribution at these temperatures. As expected, galactic winds shift the metals locked up in stars to lower stellar metallicities.

In the bottom panels of Fig.~\ref{fig-zcool0metevol} we show the
metallicities of various phases as a function of redshift. As discussed in \citet{Wiersma2009b}, in the
\reference\ model most of the metals are initially in the gas phase,
while by the present day the majority of the metals are locked up in
stars. The effect of neglecting metal cooling is small for the
metallicity of either stars or SF gas, but is obvious for the
metallicity of the IGM (which is {\em lower} by up to an order of magnitude in the absence of metal cooling) and the WHIM
({\em higher} by $\approx 0.2$ dex without metal cooling).

Without winds, more metals are in stars and significantly
fewer in NSF gas, but the metallicities of stars, SF gas, dense NSF gas, and
ICM are in fact not very different. The absence of feedback has a major
effect on the metallicity of the IGM (low $T$ and
low $\rho$ NSF gas) and the WHIM (hot, NSF gas), decreasing their
metallicities by more than 2 and 1-2 orders of magnitude, respectively. Without 
galactic winds, metals
are not efficiently transported to the lower densities of the diffuse IGM or
WHIM. A careful census of the IGM metallicity, and its evolution, is
therefore a powerful probe of the properties of galactic winds
through the ages. On the other hand, the metallicity of the ICM is hardly affected
by the presence of winds, suggesting that dynamical processes such as ram pressure stripping dominate the enrichment of this phase. 

In summary, feedback from SNe is essential for enriching the
gas outside of haloes and increase the fraction of the metals in warm-hot gas. Metal-line cooling, on the other hand, shifts the metal distribution towards lower temperatures and, provided winds are present, strongly increases the fraction of metals in the ISM. The
metallicity of the present day ICM is $\sim 10^{-1}$ solar
and is remarkably insensitive to the presence of feedback and metal cooling. This implies a minimum metallicity for the ICM, and an enrichment mechanism that is not related to galactic winds.

\subsection{Different implementations for galactic winds \label{sec-WM}}
\begin{figure*}
\includegraphics[width=168mm]{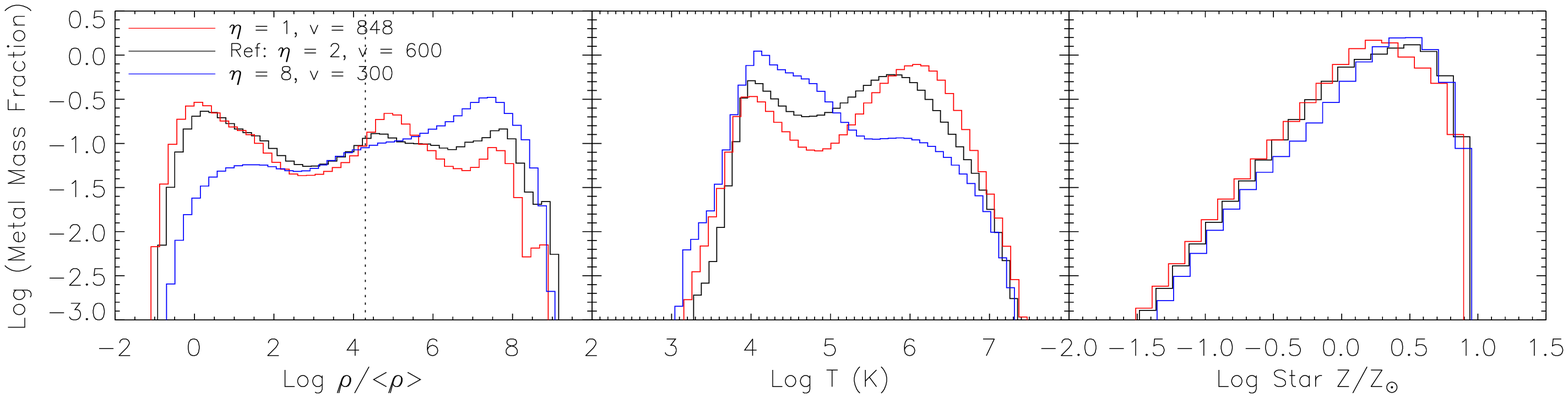}\\
\includegraphics[width=168mm]{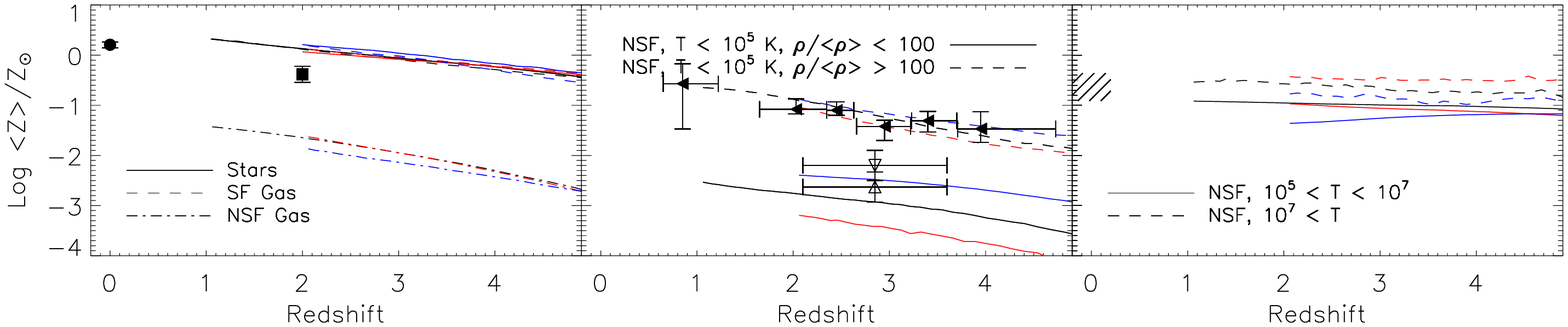}
\caption{As Fig.~\protect\ref{fig-zcool0metevol}, but for {\sc
    l025n512} simulations with different values for the mass loading,
  $\eta$, and wind speed, $v$, but identical values for $\eta v^2$ and thus for the wind energy per unit stellar mass formed. The metal mass weighted PDFs (top row) are for $z = 2$. Compared are the
  \reference\ simulation ($\eta=2$, $v_w=600~\kms$; black), a
  simulation using a mass-loading of 1 ($\eta=1$, $v_w=848~\kms$;
  red), and a simulation using a mass-loading of 8 ($\eta=8$, $v_w=300~\kms$; 
blue).  The density and temperature distributions of metals are 
  strongly dependent on $\eta$, with higher values of $\eta$ shifting metals to higher gas densities and lower temperatures.  Only the metallicity of the IGM is significantly affected, with higher $\eta$ yielding higher metallicities.}
\label{fig-windZevol}
\end{figure*}
\begin{figure*}
\includegraphics[width=168mm]{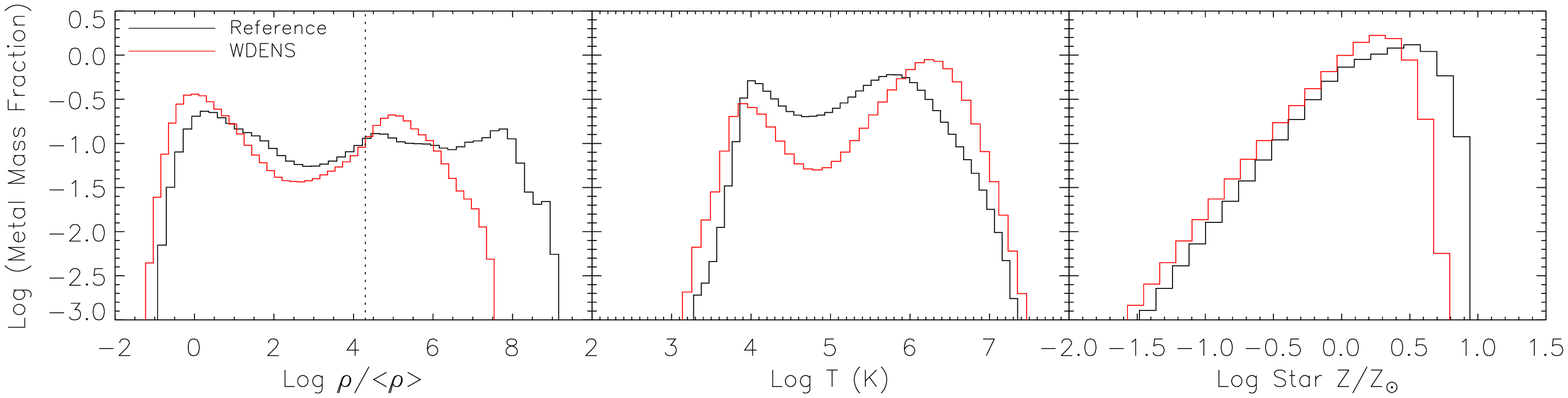}
\includegraphics[width=168mm]{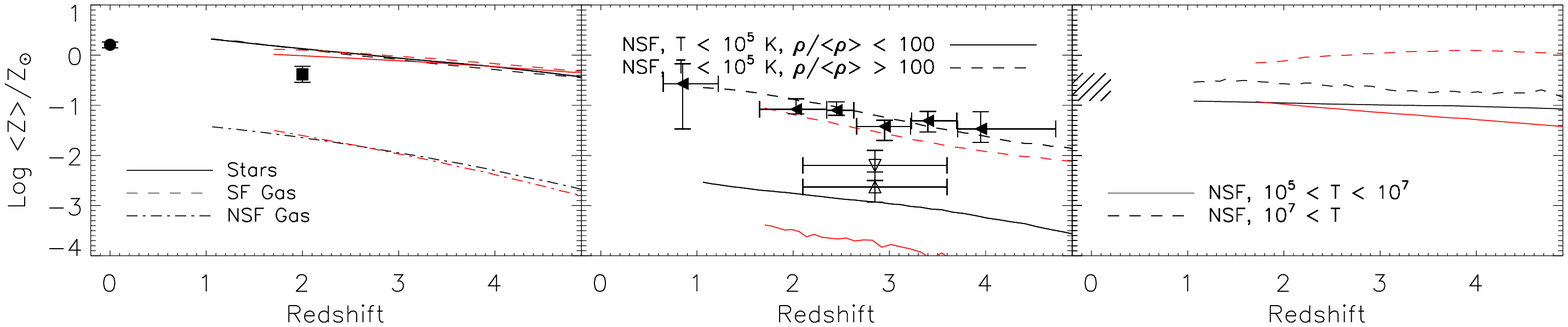}
\caption{As Fig.~\protect\ref{fig-zcool0metevol}, but for {\sc
    l025n512} simulations comparing the \reference\ simulation
  (black) to model \wdens\ (red) in which wind speed is proportional
  to the local sound speed, while keeping the the wind energy per unit stellar mass formed identical do that in the \reference\ model (Eq.~\ref{eq:wdens}). The metal mass weighted PDFs are for $z =2$. The metallicities of
  both stars and SF gas are similar between these models, but
  the top-left panel shows that lowest density gas is significantly {\em more} enriched in the
  \wdens\ model. Nevertheless, the middle bottom panel shows that the metallicity of the cold 
  IGM though is lower in the \wdens\ model, which is due to the fact that the metals are hotter in this model.}
 \label{fig-wdensZevol}
\end{figure*}
\begin{figure*}
\includegraphics[width=168mm]{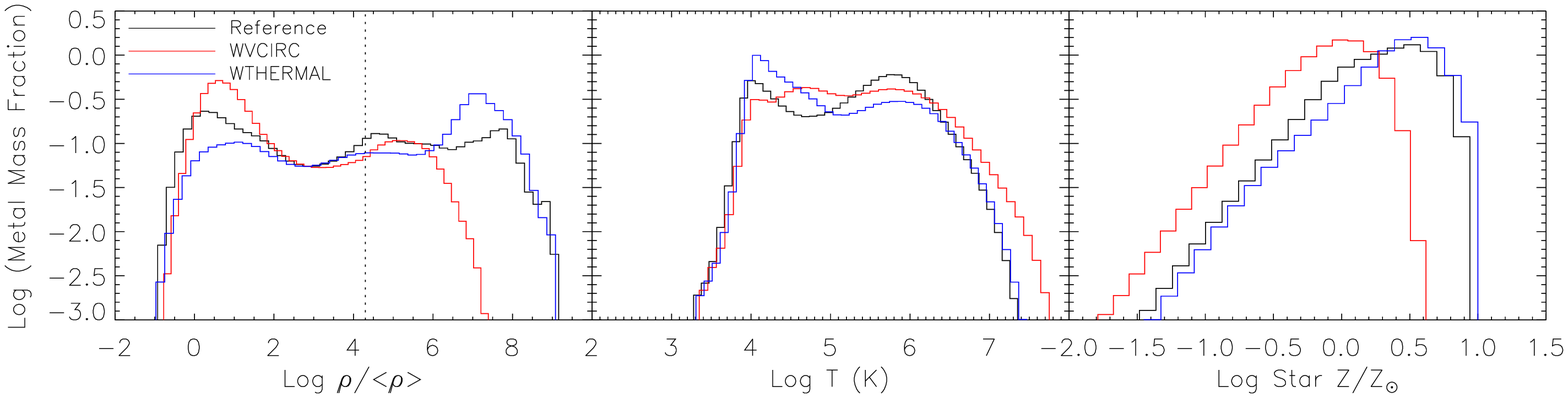}\\
\includegraphics[width=168mm]{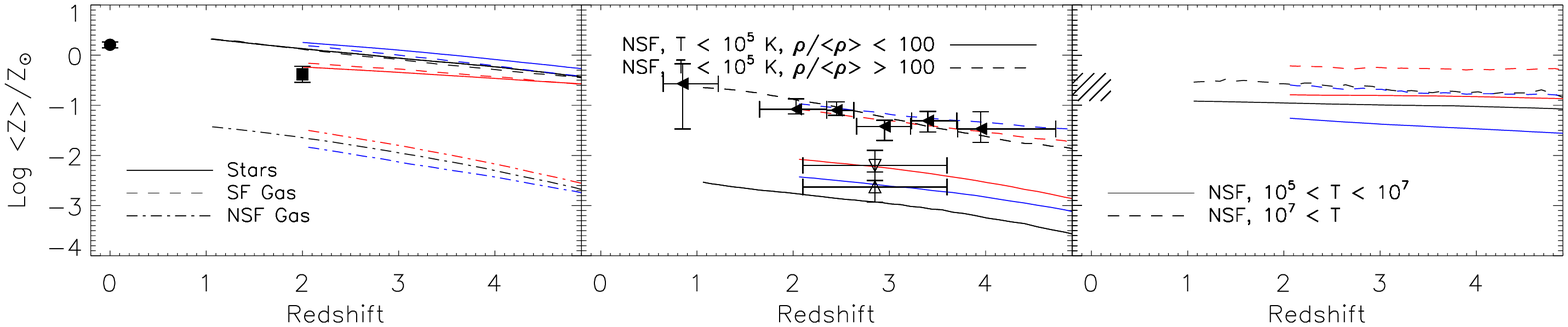}
\caption{As Fig.~\protect\ref{fig-zcool0metevol}, but for {\sc
    l025n512} simulations, comparing the \reference\ simulation
  (black) to model \wvcirc\ (red), in which wind speed depends on the
  depth of the potential well (Eq. \ref{eq:wvcirc}), 
  and to model \wthermal\, in which SN energy is
  injected thermally (blue). While \wthermal\ uses the same wind energy per unit stellar mass as the \reference\ model, the energy increases with halo mass for model \wvcirc\ (and can exceed that available from star formation). 
The metal mass weighted PDFs are for $z = 2$. 
Model \wvcirc\ 
  is able to evacuate metals from high-density regions, while at the 
  same time, spreading the metals over a variety of temperatures. Model
  \wthermal\
  differs less dramatically from the \reference\ model, with more metals 
  ending up in high-density regions.}
\label{fig-wmomthermZevol}
\end{figure*}

In the previous section we simply turned off galactic winds. Here we will investigate what happens if we vary the implementation of the outflows. We will show that the
metallicity of the IGM is particularly sensitive to the implementation of feedback, some models are quite capable of enriching this low-density gas while others cannot.  On the other hand, the metallicities 
of the stars and
the star-forming gas are largely robust to differences in the wind model. The ICM and WHIM maintain a relatively constant metallicity with redshift if the feedback recipe is varied.

Winds in most of the simulations comprising the \owls\ suite are characterised by two parameters: the
mass loading factor, $\eta$, and the initial wind speed, $v_w$
(\citealt{Dallavecchia2008}; Eq.~\ref{eq:winds}). All models shown
in Fig.~\ref{fig-windZevol} use the same value of the product
$\eta\,v_w^2$ and hence they assume that the same fraction of core
collapse SN energy is used to power winds. The wind speed sets
an approximate maximum depth of the potential well out of which the
wind can push baryons: star formation will no longer be quenched
significantly in a galaxy for which wind particles cannot escape. Note that since winds are quenched due to drag forces in the ISM, which depend on its pressure and hence on the mass of the galaxy, the relationship between
the wind speed required for escape and the potential well of the halo is non-trivial 
\citep[see][]{Dallavecchia2008}. This
parametrisation of winds is still very simple with little physical
motivation. Given the coarse resolution of cosmological simulations, wind models should be guided by observations. Unfortunately, relating the wind parameters to observations is problematic, because the wind mass loading and velocity will vary with radius (which is poorly constrained for measurements based on absorption lines), inclination, and gas phase. For a complete description of the method and further
motivation see \citet{Dallavecchia2008}.

At high $z$, when most stars form in low-mass galaxies, even a low value of $v_w$
may enable the gas to escape from most galaxies. For constant $\eta v_w^2$, a low wind speed implies high mass loading and hence a strong suppression of early star formation. Metals are then very efficiently transported out of the galaxies, and end up predominantly in the cold NSF gas, since the wind velocities are low enough that strong shocks do not occur. However, as time progresses, a low wind speed
model can no longer suppress star formation in the increasingly more
massive galaxies. The metals that are produced can no longer
escape from the haloes and remain in the SF gas. For higher values of
$v_w$ metals can escape from typical star-forming galaxies up to far lower $z$ and end up in hotter gas because the winds get shocked to higher temperatures.

The dependence of the $z=2$ metal distribution on the wind speed and mass
loading is shown in the top row of Fig.~\ref{fig-windZevol} for models that all use constant $\eta$ and $v_w$, identical $\eta v_w^2$, but for which $\eta$ varies by a factor of 8 (and thus $v_w$ by a factor of $\sqrt{8}$). 
As expected, a higher
$v_w$ results in a larger fraction of metals residing in gas with low density and high temperature. Conversely, low-$v_w$ models have a considerably
larger fraction of their metals at high density, and low temperature ($T\sim 10^4\,$K).
The low mass-loading model shows a peak in the density distribution of
the metal mass just above the star formation threshold (vertical dotted line)
that is similar to the reference model, indicating that the location of
this peak is not set by the wind velocity (although the width and the
height certainly are).  There is another small peak at 
$\rho \sim 10^{7.5} \langle \rho \rangle$ in the metal-mass weighted 
density PDF of the low mass-loading model, which may
be due to the largest haloes which have such high ISM pressures
that even these high-velocity winds are quenched. The metal mass distributions in temperature show how higher
wind velocities lead to more metals in hotter gas, due to metals in
winds getting shocked to higher temperatures. A large portion of metals that are at $T \sim
10^5\,\K$ in the \reference\ model are heated to $\ga 10^6\,\K$ for 
$v_w = 848~\kms$. The
stellar metal mass PDF shifts to slightly lower metallicities for higher $v_w$. 

While the distribution of the metals depends very strongly on the wind
model, there is relatively little difference in the {\em metallicity}
between models with different $v_w$, with the exception of the metallicity of the
IGM. In particular, the stellar
and SF gas abundances are only marginally affected by the value of $v_w$ for the range of parameter values that we investigated. The metallicities of both the WHIM and the ICM remain
strikingly
constant with time at $\approx -1$ and $\approx -0.5$, respectively,
all the way from $z=6$ to $z=1$ and, as shown in
Fig.~\ref{fig-zcool0metevol}, even to $z=0$.  These values are clearly
relatively robust with respect to the wind implementation.

The IGM metallicity in the high-mass-loading ($\eta=8$) model is nearly
an order of magnitude higher than in the low-mass-loading ($\eta=1$) model, despite 
the fact that a much smaller fraction of the metal mass resides at densities 
$\rho < 10^2~\langle \rho \rangle$ for $\eta = 8$. The increase in the 
metallicity of the IGM (which has $T < 10^5\,\K$ by definition) with decreasing 
$v_w$ must therefore be due to a decrease in the temperature of the enriched 
low-density gas. The $\eta=8$ model gives better agreement
with the value inferred for the oxygen abundance by \cite{Aguirre2008}, and is well
within the uncertainties of the \cite{Seta03} value for carbon. Because most of the metals in the IGM are ejected by
low-mass galaxies \cite[see][]{Booth2010b,Wiersma2010} from which even low-velocity winds can remove gas, a large mass-loading enriches the
low-density IGM much more efficiently.

In model \wdens\ the fraction of SN energy that powers the wind, $\epsilon_{\rm SN}\propto
\eta\,v_w^2$
 (Eq. \ref{eq:winds}),  is still kept fixed to the value used in the reference model, but $v_w$ scales with the local sound speed,
$c_s$. Since star-forming gas is on an imposed $P-\rho$ relation
(Eq. \ref{eq:EOS}), $c_s^2\propto P/\rho\propto \rho^{\gamma_{\rm eff} -1} \propto \rho^{1/3}$
for our assumed value of $\gamma_{\rm eff}=4/3$. The assumed wind
parameters therefore depend on the density:
\begin{eqnarray}
  v_{\rm w} &=& v_0 \left(\displaystyle\frac{n_{\rm H}}{n_{\rm H; thres}}\right)^{1/6} {\rm and} \\
  \eta &=& \eta_0 \left(\displaystyle\frac{n_{\rm H}}{n_{\rm H; thres}}\right)^{-2/6}
\label{eq:wdens}
\end{eqnarray}
where $n_{\rm H; thres}$ is the star formation threshold density. We
chose $v_0=600~\kms$ and $\eta_0=2$, so that the wind parameters are
the same as in the \reference\ model at the star formation
threshold. Compared with the \reference\ model, the winds in \wdens\ will remain efficient in regulating 
star formation for denser gas at the bottom of deeper potential wells, where the winds will have higher $v_w$.

Figure~\ref{fig-wdensZevol} confirms that model
\wdens\ is able to reduce the build up of metals in high-density
gas and the fraction of metals locked up in high-metallicity stars. In general, \wdens\ continues the trend with increasing $v_w$ shown in Fig.~\ref{fig-windZevol}. The high-density tail is truncated at
$\rho/\langle\rho\rangle \sim 10^7$. The higher wind
speeds result in a larger fraction of metals at $T > 10^6\,\K$ due 
to winds getting shocked. With fewer metals in high-density SF gas, 
stellar metallicities are on
average $\approx 0.1$~dex lower than in the \reference\ model.
The ICM is enriched to higher levels at nearly solar, with little
evolution. We note, however, that at these high redshifts the `ICM' consists largely of hot, low-density gas shocked by galactic winds. As we will show below (see Fig.~\ref{fig-summary5}), for $z=0$ model \wdens\ predicts ICM metallicities that are only slightly greater than for the reference model. The enrichment of the cold IGM is notably
less advanced than in the \reference\ model, with a metallicity nearly 1~dex lower, in spite of the larger fraction of metals at low densities. This is again because the faster winds shock-heat more of the escaping metals to temperatures for
which the cooling time is long. 

In model \wdens\, the wind speed scales with the local sound speed to
obtain efficient feedback in low-mass galaxies at early times\footnote{Note that lowering $v_0$, and thus increasing $\eta_0$, would further boost the efficiency of the feedback in low-mass galaxies.} as well as
in more massive haloes at later times. An alternative method to obtain
a similar behaviour makes the wind speed depend directly on the depth
of the potential well, as in the model discussed by
\citet{Oppenheimer2008}, which is meant to mimic a case where the wind
is driven by radiation pressure. Such a wind is implemented in our
\wvcirc\ model as follows: an on-the-fly friends-of-friends group finding algorithm is used to identify haloes and their circular velocities, $v_{\rm circ} = \sqrt{GM/R_{\rm vir}}$ are computed. The
wind speed and mass loading are then given by
\begin{eqnarray}
 \label{eq:wvcirc}
  v_{\rm w} &=& \displaystyle\frac{(3 + \eta_0)}{\sqrt{2}} v_{\rm circ}, \\
  \eta &=& \displaystyle\frac{v_{\rm crit} / v_{\rm circ}}{\sqrt{2}},
\end{eqnarray}
where $v_{\rm
  crit}=150~\kms$ and $\eta_0=2$ \citep{Oppenheimer2008}. 
In this model the effective value of $\epsilon_{SN}$ is no longer held constant, as the 
momentum carried by the wind is kept fixed instead. For galaxies in groups and clusters the wind energy exceeds that available from SNe. This is not necessarily a problem, as the winds may be driven by radiation pressure. However, for the parameter values of  \citet{Oppenheimer2008}, the momentum that is injected also exceeds that available in the form of radiation (Haas et al., in preparation). These caveats should be kept in mind.

Figure~\ref{fig-wmomthermZevol} shows that model \wvcirc\ transports significantly more metals to low densities, strongly reducing the fraction of metals located up in high-metallicity stars, similar to \wdens\ (Fig.~\ref{fig-wdensZevol}). However the temperature
distribution of the metals in this model is quite different from that
of \wdens\ with a much smaller fraction of metals at $\sim 10^6\,\K$ and a much larger fraction at $\sim 10^5\,\K$. \wvcirc\ predicts a
metallicity of the diffuse IGM that is up to an order of magnitude higher than for the \reference\ model. This is a consequence of the high mass loading factors used in small galaxies, which results in much more efficient transport from galaxies to the IGM, combined with the low wind velocities, which ensure that the winds do not shock the gas to high temperatures. The higher metallicity of the cold IGM agrees better with the observations, as was already shown by \citet{OD06} for a similar model. Stellar metallicities, which are a factor of $\approx 2$ lower by $z \approx 2$, also agree better with observations. 

In Fig.~\ref{fig-wmomthermZevol} we also show results from another
model, \wthermal, discussed in \cite{Schaye2010} and 
Dalla Vecchia and Schaye (in prep.), in which the energy that is injected is again identical to that in the \reference\ model, but in which this SN energy is deposited locally as {\em thermal} rather than kinetic energy. 
As star formation occurs in high density regions that cool
efficiently, such a feedback mechanism may be very inefficient if the
multi-phase nature of the ISM is not resolved \cite[e.g.][]{Katz92}.  
Our implementation of thermal feedback 
avoids much of the overcooling problem by giving the energy to only a few
particles so that they can be heated to sufficiently high temperatures that the cooling time is long \cite[see also][]{Theuns2002}. Nevertheless, model \wthermal\ predicts a greater fraction of metals to reside in dense gas and high-metallicity stars, indicating that the winds in this model are less efficient than in the \reference\ model. The metallicity of the cold IGM is, however, higher, because of the lower temperature of the enriched gas, which is no longer shock-heated by high-velocity winds. Consequently, the metallicity of the WHIM is slightly lower for model \wthermal. 

In summary, the details of the wind implementation have a large impact
on the predicted metal census. In particular, the fraction of metals
in a given phase varies strongly with the initial wind velocity and
mass loading, even if the energy in the wind is kept fixed. Compared
with the large range in metallicity spanned by the various phases, the
effects of variations in the wind implementation on the metallicities
are mostly minor. The metallicity of the cold IGM is most sensitive to
wind implementation. While models with high wind velocities, and thus
low mass loading factors, shift metals to lower densities, they tend
to predict lower metallicities for the cold IGM, as the metals are
shocked to high temperatures. Indeed, the high observed metallicities
for this phase are better matched by models with low wind
velocities. High wind velocities are, however, required for at least
the more massive galaxies to reduce the stellar metallicities to the
observed values. If we allow the wind parameters and energy to vary
with the mass of the halo, with high mass loading factors in low-mass
haloes and high wind velocities in high-mass haloes, then we can
simultaneously boost the metallicity of the cold IGM and reduce that
of the stars. This is the case for model
\wvcirc \citep{Oppenheimer2008}, although we noted that this model
uses both more momentum and energy than is available from star
formation. Finally, we note that it would be worthwhile to analyse the
gas abundances in the simulations in a way that better mimics the
observational determinations.

\subsection{Active Galactic Nuclei}
\label{sect:AGN}
\begin{figure*}
\includegraphics[width=168mm]{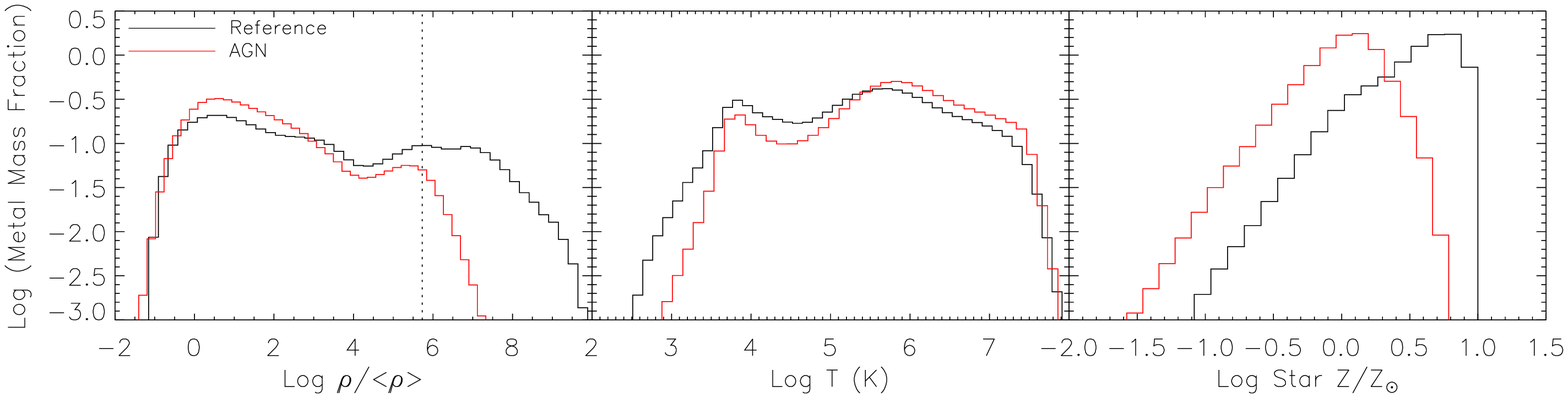}\\
\includegraphics[width=168mm]{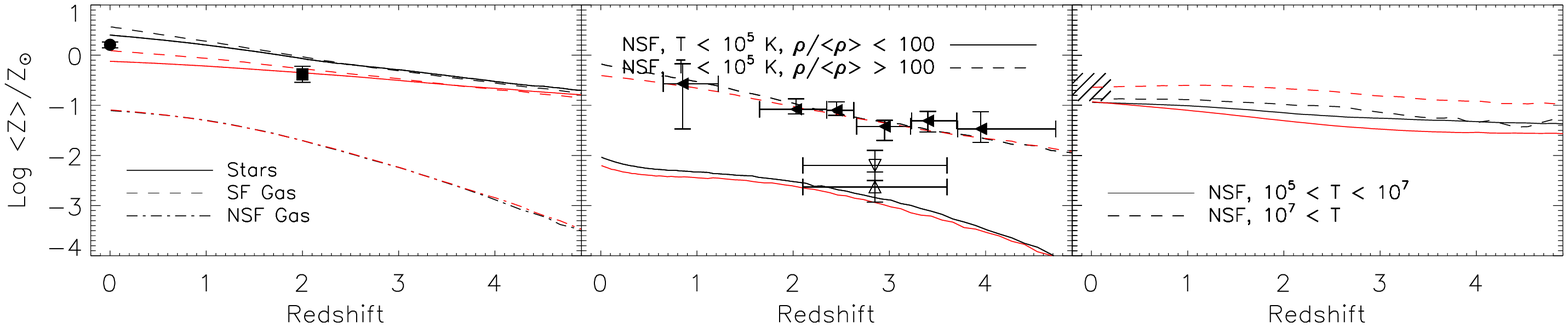}
\caption{As Fig.~\protect\ref{fig-zcool0metevol}, but for {\sc
    l100n512} simulations comparing the \reference\ simulation
  (black) to model \agn\ (red), which includes feedback from accreting black holes. 
  The metal mass weighted PDFs are for $z = 0$. \agn\ feedback is 
  able to evacuate metals from the 
  high-density regions and decreases the metallicity of stars at $z=0$ 
  by $\approx 0.5$~dex. The metallicities of most of the other phases remain 
  largely unchanged with the exception of the hottest phase whose metallicity
  increases by a about a factor of two.}
\label{fig-AGNZevol}
\end{figure*}

Feedback from accreting black holes in Active Galactic
Nuclei (AGN) has for example been invoked to reproduce the observed black hole scaling 
relations, the break in the
galaxy luminosity function, and the thermodynamic properties of groups and clusters 
of galaxies
\citep[e.g.][]{Silk1998,Kauffmann2000, DiMatteo2005, Bower06, Croton2006, Booth2009, Booth2010, McCarthy2010}. 

Our models for black hole formation and the associated AGN feedback which are 
modified versions of the model of \citet{Springel2005a} are
fully described and tested in \citet{Booth2009} (see \citealt{Schaye2010} for a comprehensive summary) and we therefore only touch on the details here. We seed low-mass haloes with central black holes, which are then allowed to grow by
mergers and gas accretion. A fixed fraction of 1.5\% of the accreted rest mass energy is injected into the ambient gas by heating it to a very high
temperature. This model reproduces the
\citet{Magorrian1998} relation, the normalisation of which is determined by the assumed efficiency factor of the feedback, as well as the other observed black hole scaling relations \citep{Booth2009,Booth2010} and both the optical and X-ray properties of observed groups of galaxies \citep{McCarthy2010}.

Fig.~\ref{fig-AGNZevol} shows that AGN feedback is able to remove
metals from high-density regions and transport them outside haloes (i.e., $\rho < 200~\langle\rho\rangle$), while heating them to high temperatures. The metallicity of the stars that contain most of the metals is reduced by nearly an order of magnitude. 
The metallicity evolution of the other phases is similar to the \reference\ model, although the metallicity of the ICM is about a factor of two higher.  Since there are less metals in the 
star-forming regions, it is not surprising that model \agn\ has lower stellar metallicities. However the WHIM's metallicity is also
slightly reduced, notwithstanding the larger fraction of metals in the
WHIM (as seen in the temperature PDF). AGN feedback 
(as implemented here) therefore increases the fraction of baryons in the WHIM. 

\subsection{A top-heavy IMF in starbursts}
\label{sec-dbimf}
\begin{figure*}
\includegraphics[width=168mm]{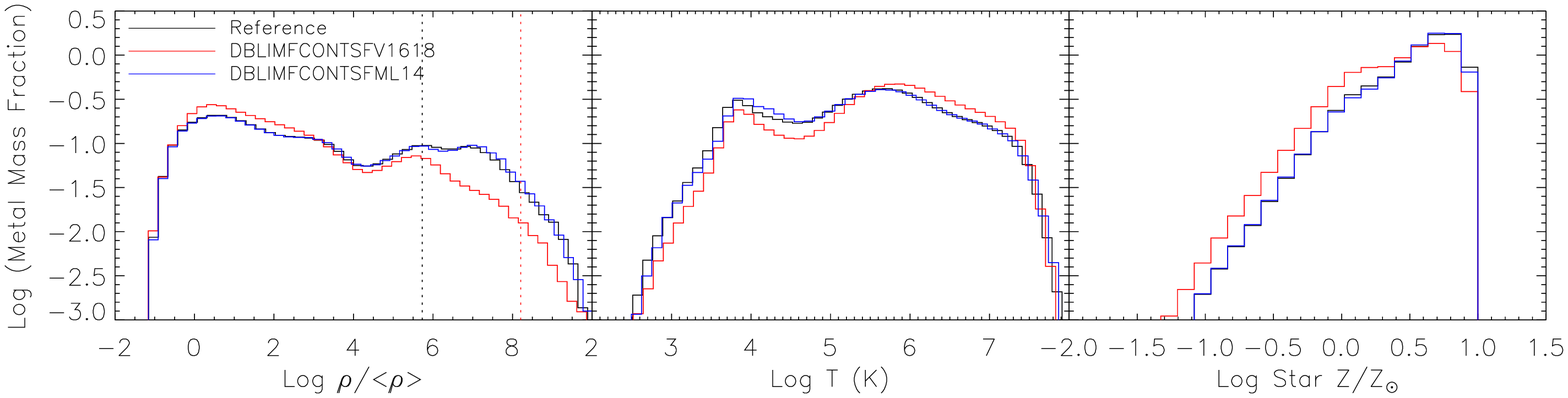}\\
\includegraphics[width=168mm]{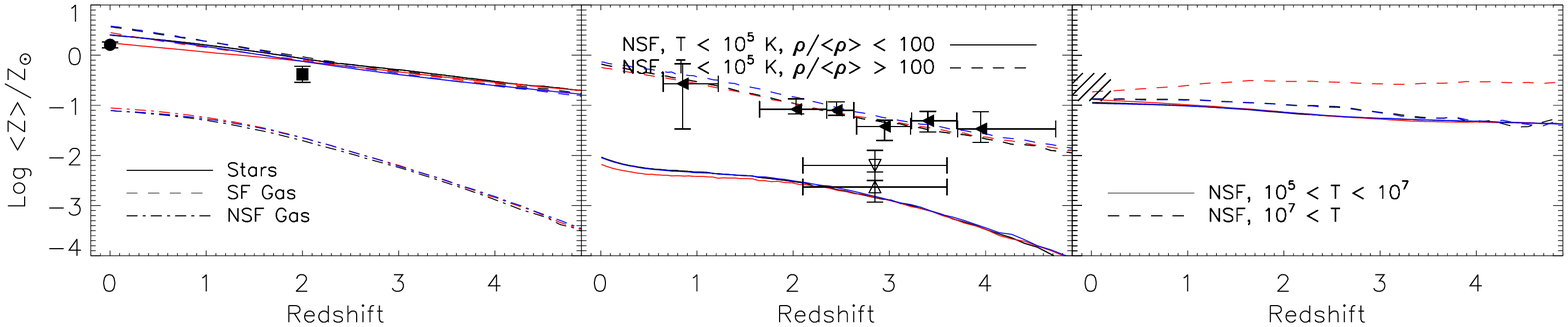}
\caption{As Fig.~\protect\ref{fig-zcool0metevol}, but for {\sc
    l100n512} simulations comparing the \reference\ simulation
  (black), with two models with a top-heavy IMF at high-pressure; one 
  in which the extra wind energy is used to increase the wind velocity 
  ($v_w = 1618~\kms$, \textsc{dblimfcontsfv1618}, red), and one in which the 
  extra wind energy is used to increase the mass loading 
  ($\eta = 14$, \textsc{dblimfcontsfml14}, blue).  The metal mass weighted PDFs
  are for $z = 0$. In the top-left panel, the black and red dotted lines indicate the star
  formation threshold and the density above which the IMF becomes top-heavy, respectively. A top-heavy IMF in starbursts only makes a difference if the extra SN energy is used to increase the wind velocity, in which case it shifts metals to lower densities and higher densities and reduces the stellar metallicity.}
\label{fig-dimfZevol}
\end{figure*}

We now consider the effect of changing the IMF on the cosmic metal 
distribution. We will show that employing 
a top-heavy IMF in high-pressure regions affects mainly the 
stellar metallicity, leaving the metallicities of the
other phases mostly unchanged. This is largely a result of the increased energy released from 
the SNe which are more prevalent for a top-heavy IMF.

There is tentative observational evidence that the IMF in some
environments is `top-heavy' as compared to the solar
neighbourhood, such as in intense starbursts \citep[e.g.]{McCrady2003}, or
the inner parsec of the Galaxy \citep[e.g.]{Maness2007}. Models of galaxy
formation hint that such a top-heavy IMF may be required to explain galaxy counts in the submm and far IR \citep[e.g.]{Lacey2008}. \cite{Padoan1997} and \cite{Larson2005}
discuss theoretical motivations for environmental dependence and evolution of the IMF, see \cite{Lacey2008} for more discussion.

We have implemented an environment-dependent IMF in our simulations,
labelled \dblimf\ in Table~\ref{tab-simset}, in which the IMF is changed
from the Chabrier fit to become top-heavy, $\Phi(M)\propto M^{-1}$ for all $M$ (compared with $\Phi(M)\propto M^{-2.3}$ for $M>1~$M$_\odot$ for Chabrier),
above a density threshold of $n_{\rm H}=30$ particles cm$^{-3}$, which 
corresponds to a pressure of $P/k = 2.0 \times 10^6 {\rm cm}^3\,\K$. These values were chosen because they result in $\sim 10$\% of the stellar mass forming with a top-heavy IMF. Because 
the star formation
rate density increases with pressure, the IMF becomes top-heavy
in regions with high star formation densities, i.e.\ `starbursts'. 
For the sake of brevity, we restrict ourselves to the case in 
which the star formation law is continuous across the `starburst' pressure  threshold (see \citealt{Schaye2010} for a full discussion of this point; see section \ref{sec-othersims} for the results of models in which this is not the case). Note that these models are likely less well converged with respect to the numerical resolution than the \reference\ model \cite[see][]{Schaye2010}.

A top-heavy IMF increases the number of core-collapse SNe per unit stellar mass. We account for this by increasing the energy per unit stellar mass that is injected into the winds. We explore two
extremes, one uses a higher mass-loading but the standard wind speed,
$(\eta=14, v_w=600~{\rm km}\,{\rm s}^{-1})$, the other uses the
standard mass loading but higher wind speed, $(\eta=2,v_w=1618~{\rm
  km}\,{\rm s}^{-1})$; note that this change is only enacted above the
density threshold for top-heavy star formation.

In Fig.~\ref{fig-dimfZevol} we compare {\sc dblimfcontsfv1618} ($\eta=2,v_w=1648~\kms$) and {\sc dblimfcontsfml14} ($\eta=14,v_w=600~\kms$) with the \reference\ model. While a top-heavy IMF increases the amount of metals 
produced per unit stellar mass in high-density regions due to the enhanced fraction of
high-mass (high yield, short lifetime) stars, the increased feedback serves to move a lot of these 
metals to high temperatures and low densities, resulting in a slightly \emph{lower} stellar metallicity, but a higher ICM metallicity for {\sc dblimfcontsfv1618}. Channeling the extra feedback energy into a higher 
mass loading factor yields metallicities and metal mass distributions 
that are nearly identical to those of the \reference\ model. This is because feedback in high-pressure gas is inefficient for such a low wind velocity. 

The use of a top-heavy IMF in \lq starbursts\rq, as implemented in the
{\sc dblimf} models, has surprisingly little effect on the relative abundance of
$\alpha$-enriched elements produced by these massive stars (not plotted). While [O/Fe] averaged over entire simulation volume is higher for a top-heavy IMF, the diffuse IGM and cold halo gas have a \emph{lower} $\alpha$ enhancement if the IMF is top heavy, because the $\alpha$ elements are deposited in hotter gas. This confirms that the differences between the \reference\ model and 
the top-heavy IMF in starbursts models are driven by the changes in the feedback rather than by the increased metal production.

\subsection{Other models}
\label{sec-othersims}

Although some processes can dramatically affect the cosmic metal
distribution, we will see in Figs.~\ref{fig-summary1}--\ref{fig-summary6} that there are a number of models that show little
difference.  It is remarkable that many drastic changes in physical
parameters and processes have little affect on the metallicity and
metal mass distribution of most, or even all, phases. We briefly comment on each of the models that
exhibit this behaviour and refer to \citet{Schaye2010} for more detailed descriptions of the models. 

\textbf{NOREION, REIONZ06, REIONZ12} - We
have also varied the reionization history of hydrogen in our simulations. 
Whereas
the \reference\ model assume H reionization at $z_{\rm r}=9$, these
models have no reionization, $z_r=6$ and $12$, respectively. We find that
so long as the heating from reionization happens, the simulations have
a very `short' memory of their thermal history. If there is no heating
from reionization, then a slightly larger fraction of the metals lie in cold gas, although the metallicity of cold, dense gas is actually a little lower. 

Aside from directly heating the mass, the reionization history determines the lowest halo mass in which galaxies are able to form. This may be especially important since \citet{Wiersma2010} and \citet{Booth2010b}
showed that the diffuse IGM is typically enriched by low-mass haloes 
at high redshift. Between $z = 9$ and $z = 0$, the virial mass of corresponding 
to $10^4\,\K$ increases from $\approx 2 \times 10^8\,\Msun$ 
to $\approx 3 \times 10^9\,\Msun$. Even the latter corresponds to only $\approx 3\times 10^2$ dark matter particles in our $25 \hMpc$ 
box. It is therefore likely that we have strongly underestimated the importance of the reionisation history for the metal content of the IGM. 

\textbf{SFAMPLx3, SFSLOPE1p75, SFTHRESZ} - The normalisation, power-law
index, and threshold gas density of the Kennicutt-Schmidt star formation law (Eq. \ref{eq:KS1}) are varied. The first  model assumes a normalisation that 
is higher by a factor of 3; the second model
has a slope of 1.75 (as opposed to 1.4); in the last model the
threshold for star formation decreases with increasing metallicity. For models with more efficient star formation, the fraction of the metals residing in the ISM is lower. This is because feedback allows star formation to self-regulate, such that the rate of energy injection is independent of the star formation law. Hence, a shorter gas consumption time scale is balanced by the ejection of gas (and metals) from galaxies, such that the star formation rate remains the same \citep{Schaye2010}. As the ISM has a higher metallicity than the gas at larger distances, this transfer of metals increases the metallicity of cold, halo gas, while leaving the metallicity of the ISM unchanged. 

\textbf{NOAGB\_NOSNIa} -  The
contribution from AGB and type Ia SNe to stellar mass loss and
feedback is ignored. \citet{Schaye2010} showed that this change reduces the $z=0$ star formation rate by about 40\%. We find that the metallicities drop by a similar factor, but that the metal mass fractions remain essentially unchanged. The abundance ratios (not shown) evolve very little and reflect solely the SNII yields ([O/Fe]$\approx 0.7$ for all redshifts and phases).

\textbf{EOS1p0, EOS1p67} - The slope, $\gamma_{\rm eff}$, of the $p\propto \rho^\gamma_{\rm eff}$ relation imposed on star-forming gas, is changed from the value 
$\gamma_{\rm eff}=4/3$ assumed in the
\reference\ model to 1 and 5/3, respectively. As was the case for the cosmic star formation history \citep{Schaye2010}, this change has no significant effect on the metal distribution.

\textbf{IMFSALP} - Assumes a
Salpeter IMF, as opposed to the Chabrier
IMF. The change
in IMF results in a small (less than a factor of two) decrease in the metallicities and [O/Fe] ratios (not plotted) of all phases, which reflects the lower rate of metal production per unit stellar mass and the lower star formation rates that result from the associated reduction in the cooling rates \citep[see][]{Schaye2010}.

\textbf{WHYDRODEC} - Wind
particles are temporarily `decoupled' from the hydrodynamics (i.e.\ they feel only gravity) as in the widely used model of
\citet{Springel2003}. This modification has a significant effect,
decreasing the metallicities of stars and the cold IGM by about a factor of two. Moreover, it roughly doubles the
fraction of metals in, but not the metallicity of, the WHIM at the expense of the fraction locked up in stars. We have 
relegated this simulation to
this overview due to space constraints, but it is an important variation. 

\textbf{MILL} - This model use the cosmological parameters of
the Millennium simulation \citep{Springel2005}, as opposed to
the WMAP3 values. The most significant difference is the larger value
of $\sigma_8$ in the \MILL\ model (0.9 vs.\ 0.74), which implies that
structure forms slightly earlier. The wind 
mass loading is doubled ($\eta=4,v_w=600~\kms $) in order to retain the match to the peak 
of the cosmic star formation history \citep{Schaye2010}. Compared with the \reference\ model, all of the curves shift to higher redshifts, but as Figs.~\ref{fig-summary1}--\ref{fig-summary6} show, even such a large change in cosmology has only a minor effect on the predictions for $z=0$ and 2.

\textbf{DBLIMFML14, DBLIMFV1618} - These
models assume a top-heavy IMF for stars formed in high-pressure regions. They are identical to the models we discussed in section~\ref{sec-dbimf}, except that the star formation law is discontinuous 
across the `starburst' threshold. Since no discontinuities in star formation 
density (as a function of gas density) have been observed, this model decreases 
the efficiency of star formation in high-pressure regions in order to produce a continuous UV flux as a 
function of pressure. The fact that these models show little difference to their 
\textsc{dblimfcontsf} counterparts is consistent with our finding that variations in the star formation law are unimportant. 

\textbf{NOHeHEAT} - As discussed in section \ref{sec-method},
we include some heating around $z \approx 3$ to account for the rise in
the IGM temperature around the time of helium reionization. Ignoring this extra heat input does not yield any significant differences, as was also true for the cosmic star formation history \citep{Schaye2010}. 

\textbf{SNIaGAUS} - Given the large uncertainties in the 
normalisation and shape of the cosmic SNIa rate, we have run a model 
where it is based on a Gaussian delay function (rather than the e-folding 
delay function used
in the \reference\ model). This results in roughly the same number of SNIa, 
but they happen much later in the history of the universe. The evolution 
of the [O/Fe] ratio shows the expected differences (the dip to lower 
values occurring at lower redshift), but the overall distribution of 
metals and metallicity are nearly identical to the results for the 
\reference\ model.

\subsection{Summary}
\label{sec-sum}

\begin{figure}
  \includegraphics[width=84mm]{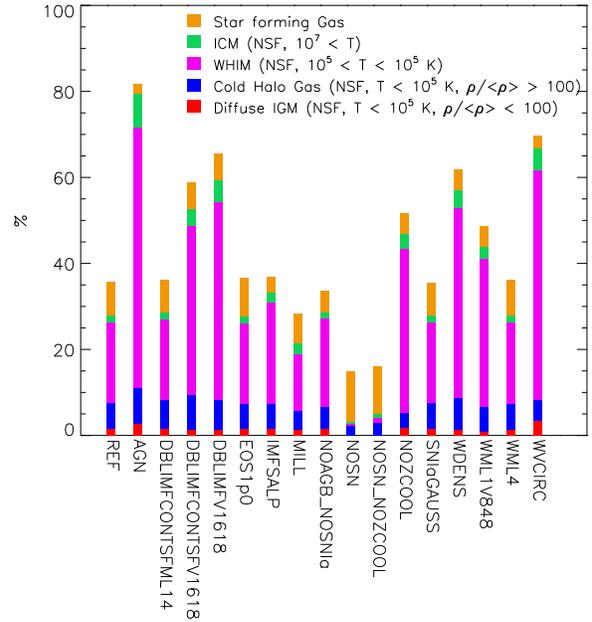}
      \caption{Fraction of metal mass in various phases for a
        selection of \textsc{l100n512} models of the \owls\ suite at $z=0$. The 
        orange, green, magenta, blue and red bars correspond to the metal mass fraction 
        in star-forming gas, ICM, WHIM, cold halo gas, and diffuse IGM, respectively. 
        The
        remainder of metals is locked up in stars. Metal fractions depend largely on the strength of feedback - models 
        without feedback have very high stellar metal fractions, whereas models with 
        strong feedback predict that most metals reside in the WHIM.}
      \label{fig-summary1}
\end{figure}

\begin{figure}
      \includegraphics[width=84mm]{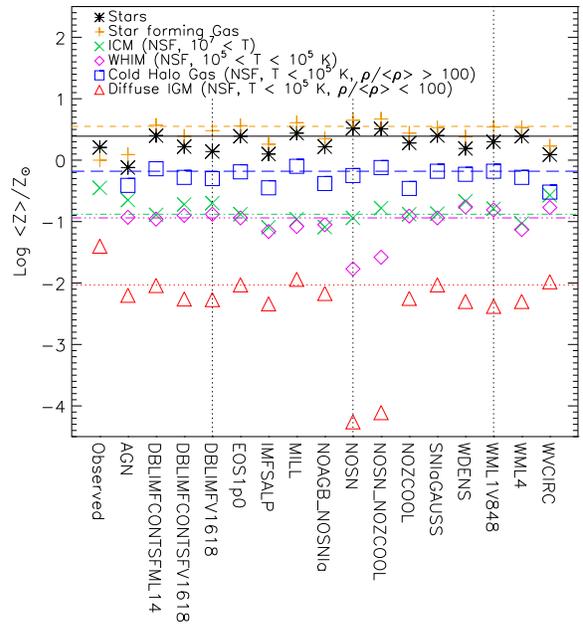}
      \caption{Metallicities of various phases for a selection of
        \textsc{l100n512} models of the \owls\ suite at $z=0$. The values for the \textsc{reference} model are indicated by the horizontal lines. Observations shown are as
        follows: stars - \protect\cite{Gallazzi2008}; ICM -
        \protect\cite{Simionescu2009}; the rest - 
        estimates compiled by \protect\cite{Pagel2008}. The
        metallicities are surprisingly similar across most models, 
        but models without
        feedback are clearly different.}
      \label{fig-summary5}
\end{figure}

\begin{figure}
  \includegraphics[width=84mm]{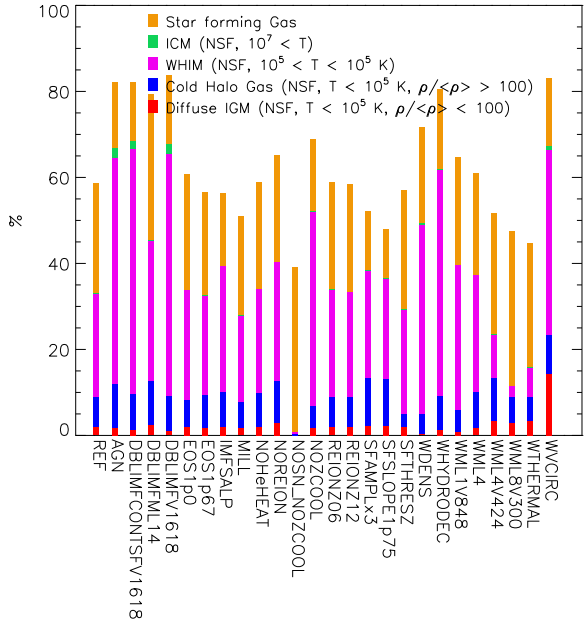}
      \caption{As Fig.~\ref{fig-summary1}, but for $z=2$.}
      \label{fig-summary2}
\end{figure}

\begin{figure}
  \includegraphics[width=84mm]{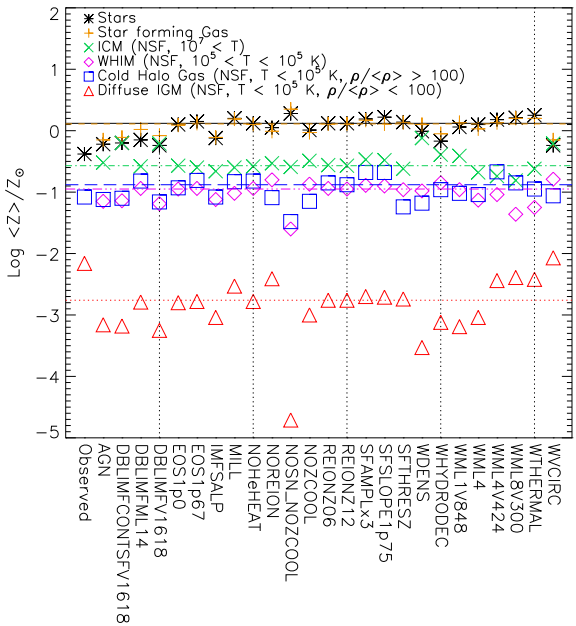}
      \caption{As Fig.~\ref{fig-summary5}, but for $z=2$. Observations shown are as follows: stars -
        \protect\cite{Halliday2008}, cold halo gas -
        \protect\cite{Prochaska2003}, and diffuse IGM -
        \protect\cite{Aguirre2008}. The diffuse IGM shows the most variation
        among the different models.}
      \label{fig-summary6}
\end{figure}

Figures \ref{fig-summary1} - \ref{fig-summary6} summarise our different models, comparing metal fractions
and metallicities at redshifts $z=0$ for the 100$h^{-1}$~Mpc and at $z=2$ for the 25$h^{-1}$~Mpc simulations. Note that many models were only run in one of the two boxes (see Table~\ref{tab-simset}). We give observational 
estimates of metallicities to compare to, but we caution that these are only meant to give 
an idea as to what values are reasonable, as the 
uncertainties are large.

At redshift $z=0$, the models that stand out most in terms of the metal mass 
fractions are \AGN, \wvcirc, 
\NOSN, and \textsc{nosn\_nozcool} (Fig.~\ref{fig-summary1}). The \NOSN\ models (no energy feedback from SNe) predict significantly higher metal fractions in stars, and also
higher stellar and ISM metallicities. Fig.~\ref{fig-summary5} shows that these models also
stand out in terms of the low metallicity of the WHIM and the diffuse IGM. Models \AGN\ and \wvcirc\ include the most efficient feedback, which result in much higher metal fractions for the WHIM, at the expense of that of the stellar component. While the stellar metallicities are also lower in these models, the WHIM has nearly the same metallicity, indicating that the change in its metal mass fraction is mostly due to a change in its baryonic content.

The data at $z=2$ are more uncertain and more incomplete (Figs.~\ref{fig-summary2} and \ref{fig-summary6}). The models that stood out at $z=0$ are also very different at this redshift. Models with a top-heavy IMF and high wind velocity, the simulations with hydrodynamically decoupled winds, and model \wdens\ are generally similar to \AGN\ and \wvcirc, although the latter stands out in terms of the high metallicity and metal fraction that it predicts for the cold IGM.

\section{Conclusions}
\label{sec-concs}

We have investigated the distribution of metals predicted by simulations taken from the OverWhelmingly Large Simulations project (\textsc{OWLS}; \citealt{Schaye2010}). We
considered different recipes and physical parameters in order to
determine what the factors are that shape the distribution
of metals as a function of the gas density, temperature, and metallicity.  This
builds on our previous work where we introduced our method and
demonstrated the effect of simulation box size and resolution
\citep{Wiersma2009b}.

Our results can be divided between statements about the effects of physical
processes on the metal mass distribution and statements about the
metal mass distribution itself. The former can be summarised as follows:

\begin{itemize}
\item Metal-line cooling increases the fraction of the metals that reside in stars and the ISM, at the expense of the fraction in highly overdense warm-hot gas. Metal-line cooling increases the metallicities of all cold phases. In particular, it increases the metallicities of the diffuse IGM and cold halo gas by about a factor of two. 

\item While other mechanisms can enrich the warm-cold IGM (e.g.\ ram
  pressure and tidal stripping), without SN feedback, its metallicity is far smaller and is strongly ruled out by observations of QSO absorption lines. 

\item Processes other than outflows driven by feedback from star formation and AGN contribute significantly to, or even dominate, the enrichment of the ICM. 

\item Even for a constant energy per unit stellar mass formed (i.e.\ a fixed feedback efficiency), the freedom given by the parameters of sub-grid implementations of galactic winds is very large. At a fixed efficiency, higher wind velocities (and thus correspondingly low mass loading factors) increase the fraction of the metals that reside in low-density ($\rho < 10^2\left <\rho\right >$) gas. However, since higher wind velocities also shift metals to higher temperatures, the metallicity of the diffuse, warm-cold IGM is in fact higher for lower wind velocities.

\item The metallicity of the warm-cold diffuse IGM is most sensitive to the implementation of galactic winds. 

\item 
By varying the parameters of the wind model with local properties, one can significantly affect the metal distribution in the IGM. For example, a model in which low-mass galaxies drive highly mass-loaded, slow winds while high-mass galaxies drive fast winds with low mass loading factors \citep[e.g.][]{Oppenheimer2008} can efficiently enrich the diffuse warm-cold IGM while still limiting the build up of metals in the stellar components of more massive galaxies. 

\item Feedback from AGN strongly reduces the fractions of metals that reside in the ISM and stars as well as the metallicities of these two components. Fast galactic winds in massive galaxies, e.g.\ resulting from a top-heavy IMF at high gas pressures, work in the same direction but have a smaller effect than AGN. Efficient feedback in high-mass galaxies only has a relatively small effect on the metallicities of the diffuse gas components.

\item Changes in the star formation law, the structure of the ISM, the cosmology, and the IMF all play a role, but are less important than the inclusion and implementation of metal-line cooling and, most importantly, outflows driven by feedback from star formation and accreting supermassive black holes. We also find a small effect for changes in the reionisation history, but for the case of the IGM this may be due to our limited resolution.

\end{itemize}

We summarise the metal mass distribution as follows:
\begin{itemize}
\item Stars and the WHIM are the dominant depositories of metals. In
  all models they contain together at least 78 \% of the metals at $z
  = 0$ and at least 53 \% at $z = 2$. At high redshift the ISM also
  contains a large fraction of the metals: at least 11 \% in all
  simulations at $z=2$. In all our models, and at both $z=0$ and $z=2$, the
  remaining components (diffuse cold-warm IGM, cold halo gas, and ICM)
  together contain less than a quarter of the metals.

\item The mean metallicities of the WHIM and ICM are nearly 
constant in time. In most models they increase slightly with time to $\sim 10^{-1}\,Z_\odot$ at $z = 0$. In contrast, the 
metallicities of the cold halo gas and the diffuse IGM, increase by more than an order of magnitude from $z=5$ to 0.

\item All our models predict, both for $z=0$ and 2, mean metallicities of $Z \sim Z_\odot$ for stars and ISM, and $10^{-1}\,Z_\odot \la Z < Z_\odot$ for the ICM and for cold halo gas. Except for the models without SN feedback, all simulations predict, both for $z=0$ and 2, mean metallicities of $Z \sim 10^{-1}\,Z_\odot$ for the WHIM and $10^{-3}\,Z_\odot \la Z \la 10^{-2}\,Z_\odot$ for the diffuse cold-warm IGM. 
\end{itemize}

The inclusion and implementation of outflows driven by feedback from
star formation and AGN is most important for predictions of the
distribution of metals. As cosmological simulations will need to
continue to use sub-grid implementations for these processes for the
foreseeable future, this implies that predicting the distribution of metals
will remain difficult for ab
initio models. However, as we have shown, provided that some feedback
is included, the mean metallicities of most different components can
be robustly predicted to order of magnitude.

The sensitivity of some of the results to feedback can also be regarded as fortunate. It gives us a useful tool to study the effects of outflows, which are an essential but very poorly understood ingredient of models of galaxy formation and evolution. The metallicity of the diffuse warm-cold IGM, which can be constrained using QSO absorption lines, is particularly sensitive to variations in the models and is therefore a very promising probe of the physics of galactic outflows.

\section*{Acknowledgments}

It is a great pleasure to thank all the members
of the OWLS team for discussions and help. The simulations presented here were run on Stella, the LOFAR
BlueGene/L system in Groningen, on the Cosmology Machine at the
Institute for Computational Cosmology in Durham as part of the Virgo
Consortium research programme, and on Darwin in Cambridge. This work
was sponsored by the National Computing Facilities Foundation (NCF) for
the use of supercomputer facilities, with financial support from the
Netherlands Organization for Scientific Research (NWO), also through a VIDI grant. This work was also supported by DFG Priority Program 1177.

\bibliographystyle{mn2e}
\bibliography{ms}

\end{document}